\documentclass[%
 twocolumn,
 amsmath, amssymb,
 aps,
 pre,
 longbibliography
]{revtex4-1}

\usepackage{graphicx}
\usepackage{dcolumn}
\usepackage{booktabs,subcaption,amsfonts,dcolumn}
\usepackage{multirow}
\usepackage{diagbox}

\usepackage{graphicx}
\usepackage[english]{babel}
\usepackage{color}   
\usepackage{bbm}
\usepackage{tabularx}
\usepackage{algorithm}
\usepackage{algpseudocode}
\usepackage{bm}
\usepackage{tikz}
\usepackage[pdfpagelabels,plainpages=false,bookmarks=true,colorlinks,linkcolor=red,urlcolor=blue,citecolor=blue]{hyperref}

\newtheorem{theorem}{Theorem}[section]

\newtheorem{lemma}[theorem]{Lemma}

\newcommand{\Ising}{\text{Ising}}

\newcommand{\unity}{\openone}

\newcommand{\ER}{E_{A_R}}
\newcommand{\EL}{E_{A_L}}
\newcommand{\EZL}{E_{Z_L}}

\newcommand{\TL}{E^{[W]}_{A_L}}
\newcommand{\TAR}{E^{[W]}_{A_R}}
\newcommand{\TR}{E^{[W]}_{Z_R}}

\newcommand{\rbra}[1]{(#1|}
\newcommand{\rket}[1]{|#1)}

\renewcommand{\vec}[1]{{\bf #1}}
\newcommand{\braket}[1]{\langle #1  \rangle}
\newcommand{\ket}[1]{| #1  \rangle}
\newcommand{\bra}[1]{\langle #1|}

\newcommand{\inc}{\text{inc}}

\newcommand{\norm}[1]{\left\lVert#1\right\rVert}

\newcommand{\abs}[1]{|#1|}

\newcommand{\GL}{\text{GL}}

\newcommand{\site}[1]{{[#1]}}

\renewcommand{\v}{\vec{v}}

\renewcommand{\H}{\mathcal{H}_\text{MPS}}
\newcommand{\PH}{\vec P\mathcal{H}_\text{MPS}}
\newcommand{\M}{\mathcal{M}_\text{MPS}}
\newcommand{\C}{\mathbb{C}}
\newcommand{\HH}{\mathcal{H}}

\newcommand{\HPsi}{\hat{H}\ket{\Psi}}

\begin{document}
\title{Time-dependent variational principle of mixed matrix product states in the thermodynamic limit}

\author{Yantao Wu}

\affiliation{
The Department of Physics, Princeton University
}
\date{\today}
\begin{abstract}
We describe a time evolution algorithm for quantum spin chains whose Hamiltonians are composed of an infinite uniform left and right bulk part, and an arbitrary finite region in between. 
The left and right bulk parts are allowed to be different from each other. 
The algorithm is based on the time-dependent variational principle (TDVP) of matrix product states.  
It is inversion-free and very simple to adapt from an existing TDVP code for finite systems.   
The importance of working in the projective Hilbert space is highlighted.
We study the quantum Ising model as a benchmark and an illustrative example. 
The spread of information after a local quench is studied in both the ballistic and the diffusive case. 
We also offer a derivation of TDVP directly from symplectic geometry. 
\end{abstract}

\pacs{Valid PACS appear here}
\maketitle
\section{Introduction}
Over the last two decades, research in quantum dynamics has benefited greatly from numerical algorithms that can simulate accurately the real-time dynamics of many-body quantum systems.  
For one-dimensional systems, two time evolution algorithms, both based on matrix product states (MPS), have proved reliable: the time evolving block decimation (TEBD) method \cite{tebd} and the time-dependent variational principle (TDVP) algorithm \cite{tdvp, finite_TDVP}. 
For translationally invariant systems, both methods can generalize to the thermodynamic limit: the iTEBD \cite{itebd} and the iTDVP \cite{iTDVP, Tangent_space}, eliminating the undesirable finite-size effects and reducing the complexity dependence of the system size from linear to constant. 
Based on locality \cite{Locality}, one expects that for systems composed of uniform left and right bulk parts and finite impurities in between, the time evolution algorithms should also have an efficient thermodynamic version.  
While it is not clear to us how this can be done for TEBD, a TDVP-based method to deal with such cases has been put forth in \cite{nonuniform_TDVP}.

After \cite{nonuniform_TDVP} was published, tangent space methods of MPS have developed significantly \cite{finite_TDVP, VUMPS, iTDVP, Tangent_space}. 
It is thus worthwhile to revisit the problem and apply these development. 
In this paper, we greatly simplify the algorithm in \cite{nonuniform_TDVP} and improve it in many ways.
While \cite{nonuniform_TDVP} only treats nearest-neighbor interactions, we will be able to treat any Hamiltonian that can be written as a matrix product operator (MPO).   
\cite{nonuniform_TDVP} also uses inverses of matrices conditioned by the MPS Schmidt coefficients, which can be very small.   
The algorithm described below will be completely inversion-free.
\cite{nonuniform_TDVP} considers only the Hamiltonians whose left and right bulk parts are the same, and the quenches which only change the finite region of impurities.       
We will allow the left and right bulks to be different and the quenches to change the bulk parts. 

The core idea of TDVP is very simple. 
The states representable by MPSs with a given bond dimension form a submanifold, $\H$, of the entire Hilbert space \cite{Geometry_MPS}.   
For a state, $\ket{\Psi(t)}$, at time $t$, the time evolution governed by its Hamiltonian $\hat{H}$ leads the state out of $\H$, i.e. $\hat{H}\ket{\Psi(t)}$ is not in the tangent space of $\H$ at $\ket{\Psi(t)}$. 
For the time evolution to stay in $\H$, the TDVP mandates to approximate $\hat{H}\ket{\Psi(t)}$ as its {\it orthogonal projection} onto the tangent space in the integration of the time evolution.
One then chooses a small time step, and integrates the projected $\hat{H} \ket{\Psi(t)}$ to obtain a trajectory in $\H$.    
The technical difficulty in applying TDVP to MPSs comes from the gauge freedom in an MPS, i.e. the same quantum state can be represented by two MPSs with very different matrix elements.   
This means that the time evolution of the quantum state does not uniquely specify how the matrix elements of an MPS should evolve. 
One thus needs to specify a gauge choice for the MPS and its tangent vectors.   

This paper is organized as follows. 
In Sec. \ref{sec:system}, we describe the system of interest and its MPS approximation. 
We will examine very carefully the gauge freedom of the MPS.
In Sec. \ref{sec:tangent}, we review some facts about the tangent space of $\H$ and provide a gauge choice for the tangent vectors.   
In Sec. \ref{sec:ortho}, we present the orthogonal projection of $\hat{H}\ket{\Psi}$. 
The derivation of the results in this section is technical, and is given in Appendix \ref{app:ortho}.  
In Sec. \ref{sec:integration}, we give an integration scheme to obtain the TDVP dynamics. 
In Sec. \ref{sec:Ising}, we study the quantum Ising model as an example.   
The speed of information spreading after a local quench is studied in both the ballistic and the diffusive case.  
In Sec. \ref{sec:discussion}, we discuss and conclude. 
For completeness, we give a derivation of the TDVP principle directly from symplectic geometry in Appendix \ref{app:symp}. 

\section{The system of interest, its MPS approximation, and gauge freedom}
\label{sec:system}
We consider an infinite quantum spin chain with a local Hilbert space of dimension $d$ on each site.   
The system has an infinite left and right bulk part, and a finite region of impurities with length $n_W$ in between. 
Let the Hamiltonian $\hat{H}$ be written as an infinite MPO with four-index MPO elements $W^{ss'}_{ab}$ with $a,b=1,\cdots, d_W$ and $s, s' = 1,\cdots,d$, where $d_W$ is the bond dimension of the MPO: 
\begin{equation}
  \begin{split}
    \hat{H} &= \sum_{\vec s,\vec s'}(... W_\site{i-1}^{s_{i-1}s'_{i-1}} W_\site{i}^{s_is'_i} W_\site{i+1}^{s_{i+1}s'_{i+1}}...)\ket{\vec s}\bra{\vec s'}
  \\
&=\dots
\begin{tikzpicture}[baseline = (X.base),every node/.style={scale=0.6},scale=.4]
\draw (0.5,1.5) -- (1,1.5); 
\draw[rounded corners] (1,2) rectangle (2,1);
\draw (1.5,1.5) node (X) {$W_A$};
\draw (2,1.5) -- (3,1.5); 
\draw[rounded corners] (3,2) rectangle (4,1);
\draw (3.5,1.5) node {$W_A$};
\draw (4,1.5) -- (5,1.5);
\draw[rounded corners] (5,2) rectangle (6,1);
\draw (5.5,1.5) node {$W_1$};
\draw (6,1.5) -- (6.5,1.5); 
\draw (1.5,1) -- (1.5,.5); 
\draw (1.5,2) -- (1.5,2.5); 
\draw (3.5,1) -- (3.5,.5); 
\draw (3.5,2) -- (3.5,2.5); 
\draw (5.5,1) -- (5.5,.5);
\draw (5.5,2) -- (5.5,2.5);
\draw (7.10,1.5) node[scale=1.25] (X) {\dots};
\draw (7.6,1.5) -- (8.1,1.5); 
\draw[rounded corners] (8.1,2) rectangle (9.25,1);
\draw (8.75,1.5) node (X) {$W_{n_W}$};
\draw (9.25,1.5) -- (10.25,1.5); 
\draw[rounded corners] (10.25,2) rectangle (11.25,1);
\draw (10.75,1.5) node (X) {$W_Z$};
\draw (11.25,1.5) -- (12.25,1.5); 
\draw[rounded corners] (12.25,2) rectangle (13.25,1);
\draw (12.75,1.5) node (X) {$W_Z$};
\draw (13.25,1.5) -- (13.75,1.5); 
\draw (8.75,1) -- (8.75,.5); 
\draw (8.75,2) -- (8.75,2.5); 
\draw (10.75,1) -- (10.75,.5); 
\draw (10.75,2) -- (10.75,2.5); 
\draw (12.75,1) -- (12.75,.5);
\draw (12.75,2) -- (12.75,2.5);
\end{tikzpicture} \dots
\end{split}
\label{eq:MPO}
\end{equation}
where $W_{[i]} = W_A$ for all lattice sites $i < 1$ and $W_{[i]} = W_Z$ for all $i > n_W$, and $W_{[i]}$ are arbitrary for $i = 1,\cdots, n_W$. 
In the following, for notational conciseness, we drop the physical index $s$ on the tensors in an MPS or an MPO when confusion does not arise. 
Based on locality principles like the Lieb-Robinson bound \cite{Lieb-Robinson}, we assume that the MPS approximating the time-evolved quantum states has the form 
\\
\begin{equation}
  \begin{split}
    |\Psi(A; &B^i; Z)\rangle = \sum_{\vec s}(... A_\site{i-1}^{s_{i-1}} A_\site{i}^{s_i} A_\site{i+1}^{s_{i+1}}...)\ket{\vec s}
\\ 
&=\dots
\begin{tikzpicture}[baseline = (X.base),every node/.style={scale=0.6},scale=.4]
\draw (0.5,1.5) -- (1,1.5); 
\draw[rounded corners] (1,2) rectangle (2,1);
\draw (1.5,1.5) node (X) {$A$};
\draw (2,1.5) -- (3,1.5); 
\draw[rounded corners] (3,2) rectangle (4,1);
\draw (3.5,1.5) node {$A$};
\draw (4,1.5) -- (5,1.5);
\draw[rounded corners] (5,2) rectangle (6,1);
\draw (5.5,1.5) node {$B^1$};
\draw (6,1.5) -- (6.5,1.5); 
\draw (1.5,1) -- (1.5,.5); 
\draw (3.5,1) -- (3.5,.5); 
\draw (5.5,1) -- (5.5,.5);
\draw (7.10,1.5) node[scale=1.25] (X) {\dots};
\draw (7.75,1.5) -- (8.25,1.5); 
\draw[rounded corners] (8.25,2) rectangle (9.25,1);
\draw (8.75,1.5) node (X) {$B^n$};
\draw (9.25,1.5) -- (10.25,1.5); 
\draw[rounded corners] (10.25,2) rectangle (11.25,1);
\draw (10.75,1.5) node (X) {$Z$};
\draw (11.25,1.5) -- (12.25,1.5); 
\draw[rounded corners] (12.25,2) rectangle (13.25,1);
\draw (12.75,1.5) node (X) {$Z$};
\draw (13.25,1.5) -- (13.75,1.5); 
\draw (8.75,1) -- (8.75,.5); 
\draw (10.75,1) -- (10.75,.5); 
\draw (12.75,1) -- (12.75,.5);
\end{tikzpicture} \dots
\end{split}
\label{eq:MPS}
\end{equation}
where $n$, the number of inhomogeneous tensors $B^i$, needs to be larger than $n_W$. 
We require $A_\site{i} = A$ for all $i < 1$, and $A_\site{i} = Z$ for all $i > n$. 
The tensors $A_\site{i}$ on lattice sites $1$ to $n$ are denoted as $B^i$ and are allowed to change arbitrarily, except restrained by the bond dimension $D$. 
In the following analysis, in order for the variational manifold to be well-defined, we fix the bond dimension of the MPS to a given value.  
Here we note that as the local information spreads with real-time dynamics in a spin chain, in order for the MPS approximation to remain accurate, $n$ needs to increase with time.  
As shown in Sec.~\ref{sec:integration}, it is very easy to expand $n$ dynamically.  
For now, we take it to be a fixed number. 

We comment here that the MPO and MPS are only well-defined for a finite system with boundary tensors at the left and the right end. 
In Eq. \ref{eq:MPO} and \ref{eq:MPS}, we have effectively taken the system size to infinity and put the boundary tensors at the left and right infinities.  
In the thermodynamic limit, the precise values of the boundary tensors do not matter, and we do not keep track of them.   
\subsection{Gauge freedom}
Eq.~\ref{eq:MPS} defines the variational manifold used to describe the time evolution of the system.    
$A$, $B^1, \cdots, B^n$, $Z$ are all complex tensors of dimension $d\times D\times D$, constituting the manifold of variational coefficients that we have access to:  
\begin{equation}
  \M = \C^{d\times D\times D} \times \C^{d\times D\times D} \times \prod_{i=1}^n \C^{d\times D\times D}. 
\end{equation}
The variational manifold of quantum states is then 
\begin{equation}
  \H = \{\ket{\Psi(A; B^i; Z)} | (A; B^i; Z) \in \M \}. 
\end{equation}
The (complex) dimension of $\M$ is much larger than that of $\H$, because of the gauge symmetries in an MPS.  
In Appendix \ref{app:ortho}, it will turn out that it is necessary to work in the projective space of $\H$:
\begin{equation}
  \PH = \H/\C,
\end{equation}
which has more gauge symmetries than $\H$. 

To quantify the MPS gauge freedom in $\PH$, we need to find the gauge group $G$ whose action on $\M$ leaves the quantum state invariant up to a scalar multiplication.  
To find $G$, first note that the following transformation leaves the quantum state invariant:    
\begin{equation}
  A' = X^{-1}_A A X_A, \,Z' = X^{-1}_Z Z X_Z, \,B'^i = X_i^{-1} B^i X_{i+1} 
    \label{eq:X}
\end{equation}
where the $X$s are arbitrary $D\times D$ invertible matrices, and $X_{1} = X_A$ and $X_{n+1} = X_Z$. 
Also note that when $X_A = X_i = X_Z = aI$, where $a$ is a complex number and $I$ is the identity matrix, the transformation in Eq. \ref{eq:X} does not change $A$, $B^i, Z$ at all, and should be excluded from the gauge group.
This means that $(\prod_{i=1}^{n+1} \GL(\C; D))/\C$ is a part of $G$, where $\GL(\C;D)$ is the multiplicative group of complex matrices of dimension $D \times D$ and $\C$ is the group of scalar multiplication. 
Because we work in the thermodynamic limit, the effect of Eq. \ref{eq:X} on the boundary tensors at the left and right infinities can be ignored. 
Because we are interested in the projective space, scalar multiplications on $A$ and $Z$ are also gauge transformations:  
\begin{equation}
  A' = \alpha A, \, Z' = \zeta Z
  \label{eq:CA}
\end{equation}
where $\alpha$ and $\zeta$ are two complex numbers. 
Scalar multiplications on $B^i$ can be accomplished by combining the transformations in Eq. \ref{eq:X} and Eq. \ref{eq:CA}. 
Thus, the full gauge group is 
\begin{equation}
  G = \C_A \times \C_Z \times (\prod_{i=1}^{n+1} \GL(\C; D))/\C 
\end{equation}
where $\C_A$ and $\C_Z$ are groups of scalar multiplication on $A$ and $Z$, each with complex dimension one. 
The complex dimension of $G$ is then the number of the complex equations that one can impose in the gauge choice of tangent vectors to $\PH$. 
It is equal to 
\begin{equation}
\dim_\C G = 1 + 1 + (n+1)D^2 - 1 = 2D^2 + (n-1)D^2 + 1. 
\end{equation}
\subsection{Mixed canonical form of MPS}
The gauge freedom of an MPS can be exploited to bring the MPS in a convenient form. 
For a entirely uniform MPS, as in the standard practice, one can write it in the mixed canonical form \cite{Tangent_space}:
\begin{align*}
\ket{\Psi(A)} &= 
\dots \begin{tikzpicture}[baseline = (X.base),every node/.style={scale=0.6},scale=.4]
\draw (0.5,0) -- (1,0);
\draw[rounded corners] (1,0.5) rectangle (2,-0.5);
\draw (1.5,0) node (X) {$A$};
\draw (2,0) -- (3,0); 
\draw[rounded corners] (3,0.5) rectangle (4,-0.5);
\draw (3.5,0) node {$A$};
\draw (4,0) -- (5,0);
\draw[rounded corners] (5,0.5) rectangle (6,-0.5);
\draw (5.5,0) node {$A$};
\draw (6,0) -- (9,0);
\draw[rounded corners] (9,0.5) rectangle (10,-0.5);
\draw (9.5,0) node {$A$};
\draw (10,0) -- (11,0);
\draw[rounded corners] (11,0.5) rectangle (12,-0.5);
\draw (11.5,0) node {$A$};
\draw (12,0) -- (12.5,0);
\draw (1.5,-.5) -- (1.5,-1); \draw (3.5,-.5) -- (3.5,-1); \draw (5.5,-.5) -- (5.5,-1); \draw (11.5,-.5) -- (11.5,-1);
\draw (9.5,-.5) -- (9.5,-1); 
\draw (5,1) node {$\phantom{X}$};
\end{tikzpicture}  \dots  \\
&= 
\dots \begin{tikzpicture}[baseline = (X.base),every node/.style={scale=0.6},scale=.4]
\draw (0.5,0) -- (1,0);
\draw[rounded corners] (1,0.5) rectangle (2,-0.5);
\draw (1.5,0) node (X) {$A_L$};
\draw (2,0) -- (3,0); 
\draw[rounded corners] (3,0.5) rectangle (4,-0.5);
\draw (3.5,0) node {$A_L$};
\draw (4,0) -- (5,0);
\draw[rounded corners] (5,0.5) rectangle (6,-0.5);
\draw (5.5,0) node {$A_C$};
\draw (6,0) -- (9,0);
\draw[rounded corners] (9,0.5) rectangle (10,-0.5);
\draw (9.5,0) node {$A_R$};
\draw (10,0) -- (11,0);
\draw[rounded corners] (11,0.5) rectangle (12,-0.5);
\draw (11.5,0) node {$A_R$};
\draw (12,0) -- (12.5,0);
\draw (1.5,-.5) -- (1.5,-1); \draw (3.5,-.5) -- (3.5,-1); \draw (5.5,-.5) -- (5.5,-1); \draw (11.5,-.5) -- (11.5,-1);
\draw (9.5,-.5) -- (9.5,-1); 
\draw (5,1) node {$\phantom{X}$};
\end{tikzpicture}  \dots  \\
&= \dots \begin{tikzpicture}[baseline = (X.base),every node/.style={scale=0.6},scale=.4]
\draw (0.5,0) -- (1,0);
\draw[rounded corners] (1,0.5) rectangle (2,-0.5);
\draw (1.5,0) node (X) {$A_L$};
\draw (2,0) -- (3,0); 
\draw[rounded corners] (3,0.5) rectangle (4,-0.5);
\draw (3.5,0) node {$A_L$};
\draw (4,0) -- (5,0);
\draw[rounded corners] (5,0.5) rectangle (6,-0.5);
\draw (5.5,0) node {$A_L$};
\draw (6,0) -- (7,0);
\draw(7.5,0) circle (0.5);
\draw (8,0) -- (9,0);
\draw (7.5,0) node {$C_A$};
\draw[rounded corners] (9,0.5) rectangle (10,-0.5);
\draw (9.5,0) node {$A_R$};
\draw (10,0) -- (11,0);
\draw[rounded corners] (11,0.5) rectangle (12,-0.5);
\draw (11.5,0) node {$A_R$};
\draw (12,0) -- (12.5,0);
\draw (1.5,-.5) -- (1.5,-1); \draw (3.5,-.5) -- (3.5,-1);  \draw (5.5,-.5) -- (5.5,-1); \draw (11.5,-.5) -- (11.5,-1);
\draw (9.5,-.5) -- (9.5,-1); 
\draw (5,1) node {$\phantom{X}$};
\end{tikzpicture}  \dots  
\end{align*}
The tensors $\{A_L, A_R, A_C, C_A\}$ satisfy the following relations:
\begin{equation}
\begin{tikzpicture}[baseline = (X.base),every node/.style={scale=0.6},scale=.4]
\draw (1,-1.5) edge[out=180,in=180] (1,1.5);
\draw[rounded corners] (1,2) rectangle (2,1);
\draw[rounded corners] (1,-1) rectangle (2,-2);
\draw (1.5,1) -- (1.5,-1);
\draw (1.5,1.5) node {$A_L$};
\draw (1.5,-1.5) node {$\bar{A}_L$};
\draw (2,1.5) -- (2.5,1.5); \draw (2,-1.5) -- (2.5,-1.5);
\end{tikzpicture} 
= 
\begin{tikzpicture}[baseline = (X.base),every node/.style={scale=0.60},scale=.4]
\draw (1,-1.5) edge[out=180,in=180] (1,1.5);
\end{tikzpicture}
\hspace{10mm}
\begin{tikzpicture}[baseline = (X.base),every node/.style={scale=0.60},scale=.4]
\draw (0.5,1.5) -- (1,1.5); \draw (0.5,-1.5) -- (1,-1.5);
\draw[rounded corners] (1,2) rectangle (2,1);
\draw[rounded corners] (1,-1) rectangle (2,-2);
\draw (1.5,1) -- (1.5,-1);
\draw (1.5,1.5) node {$A_R$};
\draw (1.5,-1.5) node {$\bar{A}_R$};
\draw (2,-1.5) edge[out=0,in=0] (2,1.5);
\end{tikzpicture} 
= 
\begin{tikzpicture}[baseline = (X.base),every node/.style={scale=0.60},scale=.4]
\draw (1,-1.5) edge[out=0,in=0] (1,1.5);
\end{tikzpicture}
\label{eq:canonical}
\end{equation}
and
\begin{equation}
\begin{split}
\begin{tikzpicture}[baseline = (X.base),every node/.style={scale=0.60},scale=.4]
\draw (0,-0.5) -- (0.5,-0.5);
\draw[rounded corners] (0.5,0) rectangle (1.5,-1);  
\draw (1,-0.5) node (X) {$A_C$};
\draw (1,-1) -- (1,-1.5);
\draw (1.5,-0.5) -- (2,-0.5);
\end{tikzpicture} 
&= 
\begin{tikzpicture}[baseline = (X.base),every node/.style={scale=0.60},scale=.4]
\draw (2.5,0) -- (3,0);
\draw[rounded corners] (3,0.5) rectangle (4,-0.5);
\draw (3.5,0) node (X) {$A_L$};
\draw (3.5,-0.5) -- (3.5,-1);
\draw (4,0) -- (5,0);
\draw (5.5,0) circle (0.5);
\draw (5.5,0) node {$C_A$};
\draw (6,0) -- (6.5,0);
\end{tikzpicture} 
= 
\begin{tikzpicture}[baseline = (X.base),every node/.style={scale=0.60},scale=.4]
 \draw (0.5,0) -- (1,0);
\draw (1.5,0) circle (0.5);
\draw (1.5,0) node (X){$C_A$};
\draw (2,0) -- (3,0);
\draw[rounded corners] (3,0.5) rectangle (4,-0.5);
\draw (3.5,0) node {$A_R$};
\draw (3.5,-0.5) -- (3.5,-1);
\draw (4,0) -- (4.5,0);
\end{tikzpicture} 
\end{split}. 
 \label{eq:AC}
\end{equation}
The tensors $A_L$ and $A_R$ are respectively called the {\it left and right canonical forms} of $A$.  
$A_C$ is called the {\it center site tensor}, and $C_A$ the {\it bond matrix}. 
When the tensors do not have uniformity at all, similar left and right canonical tensors can be found that satisfy Eq. \ref{eq:canonical} \cite{finite_TDVP}. 
The mixed-canonical form is the key to inversion-free TDVP algorithms \cite{Tangent_space}.
Motivated by this, we also write the MPS in Eq.~\ref{eq:MPS} into the mixed-canonical form: 
\begin{equation*}
  \begin{split}
    |\Psi(&A; B^i; Z)\rangle 
\\
&=\dots
\begin{tikzpicture}[baseline = (X.base),every node/.style={scale=0.6},scale=.4]
\draw (0.5,1.5) -- (1,1.5); 
\draw[rounded corners] (1,2) rectangle (2,1);
\draw (1.5,1.5) node (X) {$A$};
\draw (2,1.5) -- (3,1.5); 
\draw[rounded corners] (3,2) rectangle (4,1);
\draw (3.5,1.5) node {$A$};
\draw (4,1.5) -- (5,1.5);
\draw[rounded corners] (5,2) rectangle (6,1);
\draw (5.5,1.5) node {$B^1$};
\draw (6,1.5) -- (7,1.5); 
\draw[rounded corners] (7,2) rectangle (8,1);
\draw (7.5,1.5) node (X) {$B^2$};
\draw (8,1.5) -- (8.5,1.5); 
\draw (9.10,1.5) node[scale=1.25] (X) {\dots};
\draw (9.75,1.5) -- (10.25,1.5); 
\draw[rounded corners] (10.25,2) rectangle (11.25,1);
\draw (10.75,1.5) node (X) {$B^n$};
\draw (11.25,1.5) -- (12.25,1.5); 
\draw[rounded corners] (12.25,2) rectangle (13.25,1);
\draw (12.75,1.5) node (X) {$Z$};
\draw (13.25,1.5) -- (14.25,1.5); 
\draw[rounded corners] (14.25,2) rectangle (15.25,1);
\draw (14.75,1.5) node (X) {$Z$};
\draw (15.25,1.5) -- (15.75,1.5); 
\draw (14.75,1) -- (14.75,.5);
\draw (1.5,1) -- (1.5,.5); 
\draw (3.5,1) -- (3.5,.5); 
\draw (5.5,1) -- (5.5,.5);
\draw (7.5,1) -- (7.5,.5); 
\draw (10.75,1) -- (10.75,.5); 
\draw (12.75,1) -- (12.75,.5);
\end{tikzpicture} \dots
\\
&=\dots
\begin{tikzpicture}[baseline = (X.base),every node/.style={scale=0.6},scale=.4]
\draw (0.5,1.5) -- (1,1.5); 
\draw[rounded corners] (1,2) rectangle (2,1);
\draw (1.5,1.5) node (X) {$A_L$};
\draw (2,1.5) -- (3,1.5); 
\draw[rounded corners] (3,2) rectangle (4,1);
\draw (3.5,1.5) node {$A_C$};
\draw (4,1.5) -- (5,1.5);
\draw[rounded corners] (5,2) rectangle (6,1);
\draw (5.5,1.5) node {$B^1_R$};
\draw (6,1.5) -- (7,1.5); 
\draw[rounded corners] (7,2) rectangle (8,1);
\draw (7.5,1.5) node (X) {$B^2_R$};
\draw (8,1.5) -- (8.5,1.5); 
\draw (9.10,1.5) node[scale=1.25] (X) {\dots};
\draw (9.75,1.5) -- (10.25,1.5); 
\draw[rounded corners] (10.25,2) rectangle (11.25,1);
\draw (10.75,1.5) node (X) {$B^n_R$};
\draw (11.25,1.5) -- (12.25,1.5); 
\draw[rounded corners] (12.25,2) rectangle (13.25,1);
\draw (12.75,1.5) node (X) {$Z_R$};
\draw (13.25,1.5) -- (14.25,1.5); 
\draw[rounded corners] (14.25,2) rectangle (15.25,1);
\draw (14.75,1.5) node (X) {$Z_R$};
\draw (15.25,1.5) -- (15.75,1.5); 
\draw (14.75,1) -- (14.75,.5);
\draw (1.5,1) -- (1.5,.5); 
\draw (3.5,1) -- (3.5,.5); 
\draw (5.5,1) -- (5.5,.5);
\draw (7.5,1) -- (7.5,.5); 
\draw (10.75,1) -- (10.75,.5); 
\draw (12.75,1) -- (12.75,.5);
\end{tikzpicture} \dots
\\
&=\dots
\begin{tikzpicture}[baseline = (X.base),every node/.style={scale=0.6},scale=.4]
\draw (0.5,1.5) -- (1,1.5); 
\draw[rounded corners] (1,2) rectangle (2,1);
\draw (1.5,1.5) node (X) {$A_L$};
\draw (2,1.5) -- (3,1.5); 
\draw[rounded corners] (3,2) rectangle (4,1);
\draw (3.5,1.5) node {$A_L$};
\draw (4,1.5) -- (5,1.5);
\draw[rounded corners] (5,2) rectangle (6,1);
\draw (5.5,1.5) node {$B^1_L$};
\draw (6,1.5) -- (7,1.5); 
\draw[rounded corners] (7,2) rectangle (8,1);
\draw (7.5,1.5) node (X) {$B^2_C$};
\draw (8,1.5) -- (8.5,1.5); 
\draw (9.10,1.5) node[scale=1.25] (X) {\dots};
\draw (9.75,1.5) -- (10.25,1.5); 
\draw[rounded corners] (10.25,2) rectangle (11.25,1);
\draw (10.75,1.5) node (X) {$B^n_R$};
\draw (11.25,1.5) -- (12.25,1.5); 
\draw[rounded corners] (12.25,2) rectangle (13.25,1);
\draw (12.75,1.5) node (X) {$Z_R$};
\draw (13.25,1.5) -- (14.25,1.5); 
\draw[rounded corners] (14.25,2) rectangle (15.25,1);
\draw (14.75,1.5) node (X) {$Z_R$};
\draw (15.25,1.5) -- (15.75,1.5); 
\draw (14.75,1) -- (14.75,.5);
\draw (1.5,1) -- (1.5,.5); 
\draw (3.5,1) -- (3.5,.5); 
\draw (5.5,1) -- (5.5,.5);
\draw (7.5,1) -- (7.5,.5); 
\draw (10.75,1) -- (10.75,.5); 
\draw (12.75,1) -- (12.75,.5);
\end{tikzpicture} \dots
\\
&=\dots
\begin{tikzpicture}[baseline = (X.base),every node/.style={scale=0.6},scale=.4]
\draw (0.5,1.5) -- (1,1.5); 
\draw[rounded corners] (1,2) rectangle (2,1);
\draw (1.5,1.5) node (X) {$A_L$};
\draw (2,1.5) -- (3,1.5); 
\draw[rounded corners] (3,2) rectangle (4,1);
\draw (3.5,1.5) node {$A_L$};
\draw (4,1.5) -- (5,1.5);
\draw[rounded corners] (5,2) rectangle (6,1);
\draw (5.5,1.5) node {$B^1_L$};
\draw (6,1.5) -- (7,1.5); 
\draw[rounded corners] (7,2) rectangle (8,1);
\draw (7.5,1.5) node (X) {$B^2_L$};
\draw (8,1.5) -- (8.5,1.5); 
\draw (9.10,1.5) node[scale=1.25] (X) {\dots};
\draw (9.75,1.5) -- (10.25,1.5); 
\draw[rounded corners] (10.25,2) rectangle (11.25,1);
\draw (10.75,1.5) node (X) {$B^n_L$};
\draw (11.25,1.5) -- (12.25,1.5); 
\draw[rounded corners] (12.25,2) rectangle (13.25,1);
\draw (12.75,1.5) node (X) {$Z_C$};
\draw (13.25,1.5) -- (14.25,1.5); 
\draw[rounded corners] (14.25,2) rectangle (15.25,1);
\draw (14.75,1.5) node (X) {$Z_R$};
\draw (15.25,1.5) -- (15.75,1.5); 
\draw (14.75,1) -- (14.75,.5);
\draw (1.5,1) -- (1.5,.5); 
\draw (3.5,1) -- (3.5,.5); 
\draw (5.5,1) -- (5.5,.5);
\draw (7.5,1) -- (7.5,.5); 
\draw (10.75,1) -- (10.75,.5); 
\draw (12.75,1) -- (12.75,.5);
\end{tikzpicture} \dots
\end{split}
\label{eq:mixed_MPS}
\end{equation*}
Here $\{A_L, A_R, A_C\}$ and $\{Z_L, Z_R, Z_C\}$ are respectively the mixed canonical tensors of a uniform MPS made of $A$ and $Z$, and satisfy Eq. \ref{eq:canonical}.  
$B^1_L, \cdots, B^{n-1}_L$ and $B^2_R, \cdots, B^n_R$ also respectively satisfy the left and right canonical relations in Eq. \ref{eq:canonical}.  
However, $B_L^n$ and $B_R^1$ do not satisfy any canonical relation, because bringing them into canonical forms will destroy the uniformity of tensor $A$ and $Z$.   
This, however, as shown in Appendix \ref{app:ortho}, is not an essential difficulty.  

\section{The tangent space of matrix product states}
\label{sec:tangent}
We now analyze the tangent space to $\PH$, following \cite{Tangent_space}. 
The tangent space of $\PH$ can be obtained from the tangent space of $\H$ by identifying tangent vectors different by multiples of $\ket{\Psi}$. 
Therefore, we will still work with tangent vectors to $\H$, knowing that we can add arbitrary multiples of $\ket{\Psi}$ to the tangent vector whenever needed. 

At $\ket{\Psi(A; B^i; Z)}$, the tangent vectors to $\H$ result from infinitesimal changes on the tensor elements: $a \equiv \delta A$, $b^i\equiv \delta B^i$, and $z \equiv \delta Z$, and are given by  
\begin{equation}
  \begin{split}
  \ket{\Phi(&a; b^i; z)} \equiv \ket{\Psi(A+a; B^i+b^i; Z+z)} - \ket{\Psi(A; B^i; Z)}
  \\
  &=\sum_{i=-\infty}^{0} \dots
\begin{tikzpicture}[baseline = (X.base),every node/.style={scale=0.6},scale=.4]
\draw (0.5,1.5) -- (1,1.5); 
\draw[rounded corners] (1,2) rectangle (2,1);
\draw (1.5,1.5) node (X) {$A$};
\draw (2,1.5) -- (3,1.5); 
\draw[rounded corners] (3,2) rectangle (4,1);
\draw (3.5,1.5) node {$a$};
\draw (4,1.5) -- (5,1.5);
\draw (3.5,0.2) node {$i$}; 
\draw[rounded corners] (5,2) rectangle (6,1);
\draw (5.5,1.5) node {$A$};
\draw (6,1.5) -- (7,1.5); 
\draw[rounded corners] (7,2) rectangle (8,1);
\draw (7.5,1.5) node (X) {$A$};
\draw (8,1.5) -- (8.5,1.5); 
\draw (9.10,1.5) node[scale=1.25] (X) {\dots};
\draw (9.75,1.5) -- (10.25,1.5); 
\draw[rounded corners] (10.25,2) rectangle (11.25,1);
\draw (10.75,1.5) node (X) {$B^n$};
\draw (11.25,1.5) -- (12.25,1.5); 
\draw[rounded corners] (12.25,2) rectangle (13.25,1);
\draw (12.75,1.5) node (X) {$Z$};
\draw (13.25,1.5) -- (13.75,1.5); 
\draw (1.5,1) -- (1.5,.5); 
\draw (3.5,1) -- (3.5,.5); 
\draw (5.5,1) -- (5.5,.5);
\draw (7.5,1) -- (7.5,.5); 
\draw (10.75,1) -- (10.75,.5); 
\draw (12.75,1) -- (12.75,.5);
\end{tikzpicture} \dots
  \\
  &+\sum_{i=1}^{n} \dots
\begin{tikzpicture}[baseline = (X.base),every node/.style={scale=0.6},scale=.4]
\draw (0.5,1.5) -- (1,1.5); 
\draw[rounded corners] (1,2) rectangle (2,1);
\draw (1.5,1.5) node (X) {$A$};
\draw (2,1.5) -- (3,1.5); 
\draw[rounded corners] (3,2) rectangle (4,1);
\draw (3.5,1.5) node {$A$};
\draw (4,1.5) -- (5,1.5);
\draw[rounded corners] (5,2) rectangle (6,1);
\draw (5.5,1.5) node {$B^1$};
\draw (6,1.5) -- (6.5,1.5); 
\draw (7.1,1.5) node[scale=1.25] (X) {\dots};
\draw (7.75,1.5) -- (8.25,1.5); 
\draw[rounded corners] (8.25,2) rectangle (9.25,1);
\draw (8.75,1.5) node (X) {$b^i$};
\draw (8.75,0.2) node {$i$}; 
\draw (9.25,1.5) -- (9.75,1.5); 
\draw (10.35,1.5) node[scale=1.25] (X) {\dots};
\draw (11,1.5) -- (11.5,1.5); 
\draw[rounded corners] (11.5,2) rectangle (12.5,1);
\draw (12.00,1.5) node (X) {$B^n$};
\draw (12.5,1.5) -- (13.5,1.5); 
\draw[rounded corners] (13.5,2) rectangle (14.5,1);
\draw (14,1.5) node (X) {$Z$};
\draw (14.5,1.5) -- (15,1.5); 
\draw (1.5,1) -- (1.5,.5); 
\draw (3.5,1) -- (3.5,.5); 
\draw (5.5,1) -- (5.5,.5);
\draw (8.75,1) -- (8.75,.5); 
\draw (12.,1) -- (12.,.5); 
\draw (14.,1) -- (14.,.5);
\end{tikzpicture} \dots
\\
&+\sum_{i=n+1}^\infty 
\dots
\begin{tikzpicture}[baseline = (X.base),every node/.style={scale=0.6},scale=.4]
\draw (0.5,1.5) -- (1,1.5); 
\draw[rounded corners] (1,2) rectangle (2,1);
\draw (1.5,1.5) node (X) {$A$};
\draw (2,1.5) -- (3,1.5); 
\draw[rounded corners] (3,2) rectangle (4,1);
\draw (3.5,1.5) node {$A$};
\draw (4,1.5) -- (5,1.5);
\draw[rounded corners] (5,2) rectangle (6,1);
\draw (5.5,1.5) node {$B^1$};
\draw (6,1.5) -- (6.5,1.5); 
\draw (7.10,1.5) node[scale=1.25] (X) {\dots};
\draw (7.75,1.5) -- (8.25,1.5); 
\draw[rounded corners] (8.25,2) rectangle (9.25,1);
\draw (8.75,1.5) node (X) {$Z$};
\draw (9.25,1.5) -- (10.25,1.5); 
\draw[rounded corners] (10.25,2) rectangle (11.25,1);
\draw (10.75,1.5) node (X) {$z$};
\draw (10.75,0.2) node (X) {$i$};
\draw (11.25,1.5) -- (12.25,1.5); 
\draw[rounded corners] (12.25,2) rectangle (13.25,1);
\draw (12.75,1.5) node (X) {$Z$};
\draw (13.25,1.5) -- (13.75,1.5); 
\draw (1.5,1) -- (1.5,.5); 
\draw (3.5,1) -- (3.5,.5); 
\draw (5.5,1) -- (5.5,.5);
\draw (8.75,1) -- (8.75,.5); 
\draw (10.75,1) -- (10.75,.5);
\draw (12.75,1) -- (12.75,.5);
\end{tikzpicture} \dots
  \\
  &=\sum_{i=-\infty}^{0} \dots
\begin{tikzpicture}[baseline = (X.base),every node/.style={scale=0.6},scale=.4]
\draw (0.5,1.5) -- (1,1.5); 
\draw[rounded corners] (1,2) rectangle (2,1);
\draw (1.5,1.5) node (X) {$A_L$};
\draw (2,1.5) -- (3,1.5); 
\draw[rounded corners] (3,2) rectangle (4,1);
\draw (3.5,1.5) node {$a_L$};
\draw (4,1.5) -- (5,1.5);
\draw (3.5,0.2) node {$i$}; 
\draw[rounded corners] (5,2) rectangle (6,1);
\draw (5.5,1.5) node {$A_R$};
\draw (6,1.5) -- (7,1.5); 
\draw[rounded corners] (7,2) rectangle (8,1);
\draw (7.5,1.5) node (X) {$A_R$};
\draw (8,1.5) -- (8.5,1.5); 
\draw (9.10,1.5) node[scale=1.25] (X) {\dots};
\draw (9.75,1.5) -- (10.25,1.5); 
\draw[rounded corners] (10.25,2) rectangle (11.25,1);
\draw (10.75,1.5) node (X) {$B^n_R$};
\draw (11.25,1.5) -- (12.25,1.5); 
\draw[rounded corners] (12.25,2) rectangle (13.25,1);
\draw (12.75,1.5) node (X) {$Z_R$};
\draw (13.25,1.5) -- (13.75,1.5); 
\draw (1.5,1) -- (1.5,.5); 
\draw (3.5,1) -- (3.5,.5); 
\draw (5.5,1) -- (5.5,.5);
\draw (7.5,1) -- (7.5,.5); 
\draw (10.75,1) -- (10.75,.5); 
\draw (12.75,1) -- (12.75,.5);
\end{tikzpicture} \dots
  \\
  &+\sum_{i=1}^{n} \dots
\begin{tikzpicture}[baseline = (X.base),every node/.style={scale=0.6},scale=.4]
\draw (0.5,1.5) -- (1,1.5); 
\draw[rounded corners] (1,2) rectangle (2,1);
\draw (1.5,1.5) node (X) {$A_L$};
\draw (2,1.5) -- (3,1.5); 
\draw[rounded corners] (3,2) rectangle (4,1);
\draw (3.5,1.5) node {$A_L$};
\draw (4,1.5) -- (5,1.5);
\draw[rounded corners] (5,2) rectangle (6,1);
\draw (5.5,1.5) node {$B^1_L$};
\draw (6,1.5) -- (6.5,1.5); 
\draw (7.1,1.5) node[scale=1.25] (X) {\dots};
\draw (7.75,1.5) -- (8.25,1.5); 
\draw[rounded corners] (8.25,2) rectangle (9.25,1);
\draw (8.75,1.5) node (X) {$b^i_L$};
\draw (8.75,0.2) node {$i$}; 
\draw (9.25,1.5) -- (9.75,1.5); 
\draw (10.35,1.5) node[scale=1.25] (X) {\dots};
\draw (11,1.5) -- (11.5,1.5); 
\draw[rounded corners] (11.5,2) rectangle (12.5,1);
\draw (12.00,1.5) node (X) {$B^n_R$};
\draw (12.5,1.5) -- (13.5,1.5); 
\draw[rounded corners] (13.5,2) rectangle (14.5,1);
\draw (14,1.5) node (X) {$Z_R$};
\draw (14.5,1.5) -- (15,1.5); 
\draw (1.5,1) -- (1.5,.5); 
\draw (3.5,1) -- (3.5,.5); 
\draw (5.5,1) -- (5.5,.5);
\draw (8.75,1) -- (8.75,.5); 
\draw (12.,1) -- (12.,.5); 
\draw (14.,1) -- (14.,.5);
\end{tikzpicture} \dots
\\
&+\sum_{i=n+1}^\infty 
\dots
\begin{tikzpicture}[baseline = (X.base),every node/.style={scale=0.6},scale=.4]
\draw (0.5,1.5) -- (1,1.5); 
\draw[rounded corners] (1,2) rectangle (2,1);
\draw (1.5,1.5) node (X) {$A_L$};
\draw (2,1.5) -- (3,1.5); 
\draw[rounded corners] (3,2) rectangle (4,1);
\draw (3.5,1.5) node {$A_L$};
\draw (4,1.5) -- (5,1.5);
\draw[rounded corners] (5,2) rectangle (6,1);
\draw (5.5,1.5) node {$B^1_L$};
\draw (6,1.5) -- (6.5,1.5); 
\draw (7.10,1.5) node[scale=1.25] (X) {\dots};
\draw (7.75,1.5) -- (8.25,1.5); 
\draw[rounded corners] (8.25,2) rectangle (9.25,1);
\draw (8.75,1.5) node (X) {$Z_L$};
\draw (9.25,1.5) -- (10.25,1.5); 
\draw[rounded corners] (10.25,2) rectangle (11.25,1);
\draw (10.75,1.5) node (X) {$z_R$};
\draw (10.75,0.2) node (X) {$i$};
\draw (11.25,1.5) -- (12.25,1.5); 
\draw[rounded corners] (12.25,2) rectangle (13.25,1);
\draw (12.75,1.5) node (X) {$Z_R$};
\draw (13.25,1.5) -- (13.75,1.5); 
\draw (1.5,1) -- (1.5,.5); 
\draw (3.5,1) -- (3.5,.5); 
\draw (5.5,1) -- (5.5,.5);
\draw (8.75,1) -- (8.75,.5); 
\draw (10.75,1) -- (10.75,.5);
\draw (12.75,1) -- (12.75,.5);
\end{tikzpicture} \dots
\end{split}
\label{eq:Phi}
\end{equation}
where we have also written $\ket{\Phi}$ in the mixed canonical form.  
The meaning of the subscripts on $a_L$, $b_L^i$, and $z_R$ will become clear in Eq. \ref{eq:V}. 
\subsection{Gauge choices of the tangent vectors}
Due to the gauge freedom, parameters $a_L, b_L^i$, and $z_R$ are redundant in describing a tangent vector to $\PH$, which poses a problem to computing the projection of $\hat{H}\ket{\Psi}$.  
We now use the gauge symmetries contained in $G$ to fix these redundancies.   
Out of the $2D^2+(n-1)D^2 + 1$ gauge symmetries of $\PH$, we impose at once $2D^2+(n-1)D^2$ restraints on $a_L, b_L^i$, and $z_R$:  
\begin{equation}
\begin{tikzpicture}[baseline = (X.base),every node/.style={scale=0.6},scale=.4]
\draw (1,-1.5) edge[out=180,in=180] (1,1.5);
\draw[rounded corners] (1,2) rectangle (2,1);
\draw[rounded corners] (1,-1) rectangle (2,-2);
\draw (1.5,1) -- (1.5,-1);
\draw (1.5,1.5) node {$a_L$};
\draw (1.5,0) node(X) {};
\draw (1.5,-1.5) node {$\bar{A}_L$};
\draw (2,1.5) -- (2.5,1.5); 
\draw (2,-1.5) -- (2.5,-1.5);
\end{tikzpicture} 
= 
0
\hspace{10mm}
\begin{tikzpicture}[baseline = (X.base),every node/.style={scale=0.6},scale=.4]
\draw (1,-1.5) edge[out=180,in=180] (1,1.5);
\draw[rounded corners] (1,2) rectangle (2,1);
\draw[rounded corners] (1,-1) rectangle (2,-2);
\draw (1.5,1) -- (1.5,-1);
\draw (1.5,1.5) node {$b^i_L$};
\draw (1.5,0) node(X) {};
\draw (1.5,-1.5) node {$\bar{B}^i_L$};
\draw (2,1.5) -- (2.5,1.5); 
\draw (2,-1.5) -- (2.5,-1.5);
\end{tikzpicture} 
= 
0
\hspace{10mm}
\begin{tikzpicture}[baseline = (X.base),every node/.style={scale=0.60},scale=.4]
\draw (0.5,1.5) -- (1,1.5); \draw (0.5,-1.5) -- (1,-1.5);
\draw[rounded corners] (1,2) rectangle (2,1);
\draw[rounded corners] (1,-1) rectangle (2,-2);
\draw (1.5,1) -- (1.5,-1);
\draw (1.5,1.5) node {$z_R$};
\draw (1.5,0) node(X) {};
\draw (1.5,-1.5) node {$\bar{Z}_R$};
\draw (2,-1.5) edge[out=0,in=0] (2,1.5);
\end{tikzpicture} 
=0 
\label{eq:tangent_canonical}
\end{equation}
where the $i$ above only goes from $1$ to $n-1$. 
We still have one last symmetry to use, which we reserve for $b_L^n$ until Eq. \ref{eq:bn_L}.  
Eq. \ref{eq:tangent_canonical} can be explicitly satisfied by giving $a_L, b_L^i,$ and $z_R$ an effective parametrization:  
\begin{equation}
\begin{split}
\begin{tikzpicture}[baseline = (X.base),every node/.style={scale=0.60},scale=.4]
\draw (-0.5,-0.5) -- (0,-0.5);
\draw[rounded corners] (0,0) rectangle (1,-1);  
\draw (0.5,-0.5) node (X) {$a_L$};
\draw (0.5,-1) -- (0.5,-1.5);
\draw (1,-0.5) -- (1.5,-0.5);
\end{tikzpicture} 
&= 
\begin{tikzpicture}[baseline = (X.base),every node/.style={scale=0.60},scale=.4]
\draw (-0.5,0) -- (0,0);
\draw[rounded corners] (0,0.5) rectangle (1,-0.5);
\draw (0.5,0) node (X) {$V_{A_L}$};
\draw (0.5,-0.5) -- (0.5,-1);
\draw (1,0) -- (2,0);
\draw[rounded corners] (2,0.5) rectangle (3.0,-0.5);
\draw (2.5,0) node {$X_{A}$};
\draw (3.0,0) -- (3.5,0);
\end{tikzpicture},
\hspace{5mm}
\begin{tikzpicture}[baseline = (X.base),every node/.style={scale=0.60},scale=.4]
\draw (-0.5,-0.5) -- (0,-0.5);
\draw[rounded corners] (0,0) rectangle (1,-1);  
\draw (0.5,-0.5) node (X) {$b^i_L$};
\draw (0.5,-1) -- (0.5,-1.5);
\draw (1,-0.5) -- (1.5,-0.5);
\end{tikzpicture} 
= 
\begin{tikzpicture}[baseline = (X.base),every node/.style={scale=0.60},scale=.4]
\draw (-0.5,0) -- (0,0);
\draw[rounded corners] (0,0.5) rectangle (1,-0.5);
\draw (0.5,0) node (X) {$V_{B^i_L}$};
\draw (0.5,-0.5) -- (0.5,-1);
\draw (1,0) -- (2,0);
\draw[rounded corners] (2,0.5) rectangle (3.0,-0.5);
\draw (2.5,0) node {$X_{B^i}$};
\draw (3.0,0) -- (3.5,0);
\end{tikzpicture} 
\\ 
\begin{tikzpicture}[baseline = (X.base),every node/.style={scale=0.60},scale=.4]
\draw (0,-0.5) -- (0.5,-0.5);
\draw[rounded corners] (0.5,0) rectangle (1.5,-1);  
\draw (1,-0.5) node (X) {$z_R$};
\draw (1,-1) -- (1,-1.5);
\draw (1.5,-0.5) -- (2,-0.5);
\end{tikzpicture} 
&=
\begin{tikzpicture}[baseline = (X.base),every node/.style={scale=0.60},scale=.4]
 \draw (0.5,0) -- (1,0);
\draw[rounded corners] (1,0.5) rectangle (2,-0.5);
\draw (1.5,0) node (X){$X_Z$};
\draw (2,0) -- (3,0);
\draw[rounded corners] (3,0.5) rectangle (4,-0.5);
\draw (3.5,0) node {$V_{Z_R}$};
\draw (3.5,-0.5) -- (3.5,-1);
\draw (4,0) -- (4.5,0);
\end{tikzpicture} 
\end{split}
\label{eq:XY}
\end{equation}
where the right (left) index of $V_{A_L}(V_{Z_R})$ has dimension $D(d-1)$. 
$V_{A_L}$ is determined by requiring its column vectors be orthonormal among themselves and orthogonal to those of $A_L$: 
\begin{equation}
\begin{tikzpicture}[baseline = (X.base),every node/.style={scale=0.6},scale=.4]
\draw (1,-1.5) edge[out=180,in=180] (1,1.5);
\draw[rounded corners] (1,2) rectangle (2,1);
\draw[rounded corners] (1,-1) rectangle (2,-2);
\draw (1.5,1) -- (1.5,-1);
\draw (1.5,1.5) node {$V_{A_L}$};
\draw (1.5,-1.5) node {$\bar{V}_{A_L}$};
\draw (2,1.5) -- (2.5,1.5); \draw (2,-1.5) -- (2.5,-1.5);
\end{tikzpicture} 
= 
\begin{tikzpicture}[baseline = (X.base),every node/.style={scale=0.60},scale=.4]
\draw (1,-1.5) edge[out=180,in=180] (1,1.5);
\end{tikzpicture} 
\hspace{10mm}
\begin{tikzpicture}[baseline = (X.base),every node/.style={scale=0.6},scale=.4]
\draw (1,-1.5) edge[out=180,in=180] (1,1.5);
\draw[rounded corners] (1,2) rectangle (2,1);
\draw[rounded corners] (1,-1) rectangle (2,-2);
\draw (1.5,1) -- (1.5,-1);
\draw (1.5,1.5) node {$A_C$};
\draw (1.5,-1.5) node {$\bar{V}_{A_L}$};
\draw (2,1.5) -- (2.5,1.5); \draw (2,-1.5) -- (2.5,-1.5);
\end{tikzpicture} 
=
\begin{tikzpicture}[baseline = (X.base),every node/.style={scale=0.6},scale=.4]
\draw (1,-1.5) edge[out=180,in=180] (1,1.5);
\draw[rounded corners] (1,2) rectangle (2,1);
\draw[rounded corners] (1,-1) rectangle (2,-2);
\draw (1.5,1) -- (1.5,-1);
\draw (1.5,1.5) node {$A_L$};
\draw (1.5,-1.5) node {$\bar{V}_{A_L}$};
\draw (2,1.5) -- (2.5,1.5); \draw (2,-1.5) -- (2.5,-1.5);
\end{tikzpicture} 
=0 .
\label{eq:V}
\end{equation}
$V_{B_L^i}$ are similarly determined for $i=1,\cdots, n-1$, and $V_{Z_R}$ is determined from a right version of Eq. \ref{eq:V}. 
A tangent vector to $\PH$ is thus given by the effective parameters $X_A$, $X_{B^i}$, $X_Z$, and $b_L^n$, where $i = 1, \cdots, n-1$. 
\section{Orthogonal projection of $\hat{H}\ket{\Psi}$} 
\label{sec:ortho}
To carry out the TDVP algorithm, one needs the orthogonal projection of $\HPsi$ on the tangent space of $\PH$ at $\ket{\Psi}$, which we denote as $\ket{\Phi(X_A;X_{B^i};X_Z;b_L^n)}_H$.     
The derivation leading to $\ket{\Phi}_H$ is technical, which we give in Appendix \ref{app:ortho}. 
Only the result is presented here. 

Before we proceed, we need some facts about the MPO transfer matrix, which for $A_L$ is defined as  
\begin{equation}
  \TL = 
  \begin{tikzpicture}[baseline = (X.base),every node/.style={scale=0.6},scale=.4]
\draw (0.5, 1.5) -- (1,1.5);
\draw (0.5, -1.5) -- (1,-1.5);
\draw (0.5, 0) -- (1,0);
\draw[rounded corners] (1,2) rectangle (2,1);
\draw[rounded corners] (1,-1) rectangle (2,-2);
\draw (1.5,1) -- (1.5,0.5);
\draw (1.5,1.5) node {$A_L$};
\draw[rounded corners] (1,-0.5) rectangle (2,0.5);
\draw (1.5,-1) -- (1.5,-0.5);
\draw (1.5,0) node(X) {$W_A$};
\draw (1.5,-1.5) node {$\bar{A}_L$};
\draw (2,1.5) -- (2.5,1.5); 
\draw (2,0) -- (2.5,0); 
\draw (2,-1.5) -- (2.5,-1.5);
\end{tikzpicture}.
\label{eq:EW}
\end{equation}
Similar MPO transfer matrices can be defined for other MPS and MPO tensors analogously. 
For a uniform MPS of tensor $A$ with $m$ sites, $\braket{\Psi|\hat{H}|\Psi} \sim (\TL)^m$ up to some unimportant boundary terms. 
The extensivity of energy thus requires that $(\TL)^m$ be asymptotically linear in $m$.    
This can only happen if the leading eigenvalue of $\TL$ equals one and is defective. 
In fact, for a typical MPO, the leading eigenvalue of $\TL$ is indeed one with algebraic multiplicity two and geometric multiplicity one \cite{VUMPS}, i.e. $\TL$ has one eigenvector and one generalized eigenvector in the leading eigenspace. 
This behavior can be attributed to the Schur form (lower triangular form) of the $W$ matrix of an MPO \cite{Schur, VUMPS}, on which we give a review in Appendix \ref{app:schur}.   
We denote the left generalized eigenvector of $\TL$ by $\rbra{L_A^\site{W}}$, and the right generalized eigenvector of $\TAR$ by $\rket{R_A^\site{W}}$.  
The $\rbra{L_A^\site{W}}$ and $\rket{R_A^\site{W}}$ can be efficiently computed by an algorithm given in the Appendix of \cite{VUMPS}. (They are known as quasi-fixed points there.)
We analogously define $\rbra{L_Z^\site{W}}$ and $\rket{R_Z^\site{W}}$. 

We now give the effective parameters, $X_A, X_{B^i}, X_Z$, and $b_L^n$, of $\ket{\Phi}_H$: 
\begin{equation}
\begin{tikzpicture}[baseline = (X.base),every node/.style={scale=0.60},scale=.4]
\draw (-0.25,-0.5) -- (0.5,-0.5);
\draw[rounded corners] (0.5,0) rectangle (1.5,-1);  
\draw (1,-0.5) node (X) {$X_A$};
\draw (1.5,-0.5) -- (2.25,-0.5);
\end{tikzpicture} 
 =
 \begin{tikzpicture}[baseline = (X.base),every node/.style={scale=0.6},scale=.4]
\draw[rounded corners] (3, 4) rectangle (4.2,-1);
\draw (3.6,1.5) node {$L_A^\site{W}$};
\draw (4.2,3.5)  -- (5.2,3.5);
\draw (4.2,1.5)  -- (5.2,1.5);
\draw (4.2,-0.5) -- (5.2,-0.5);
\draw[rounded corners] (5.2,4) rectangle (6.2,3);
\draw (5.7,3.5) node {$A_C$};
\draw (5.7,3.0) -- (5.7,2.0); 
\draw[rounded corners] (5.2,2.0) rectangle (6.2,1);
\draw (5.7,1.5) node(X) {$W_A$};
\draw (5.7,1.0) -- (5.7,-0.); 
\draw[rounded corners] (5.2,0.0) rectangle (6.2,-1);
\draw (5.7,-0.5) node {$\bar{V}_{A_L}$};
\draw (6.2,1.5) -- (7.7,1.5); 
\draw (6.2,3.5) -- (7.7,3.5); 
\draw (6.2,-.5) -- (6.7,-.5); 
edge\draw (6.7,-0.5) edge[out=0,in=0] (6.7,-1.5); 
\draw (7.7,-0.5) edge[out=180,in=180] (7.7,-1.5); 
\draw (7.7,1.5) -- (8.7,1.5); 
\draw (7.7,3.5) -- (8.7,3.5); 
\draw (7.7,-.5) -- (8.7,-.5); 
\draw[rounded corners] (8.7, 4) rectangle (9.9,-1);
\draw (9.3,1.5) node {$R_{A}^\site{W}$};
\end{tikzpicture}. 
\label{eq:X_A_final}
\end{equation}
\begin{equation}
\begin{tikzpicture}[baseline = (X.base),every node/.style={scale=0.60},scale=.4]
\draw (-0.25,-0.5) -- (0.5,-0.5);
\draw[rounded corners] (0.5,0) rectangle (1.5,-1);  
\draw (1,-0.5) node (X) {$X_Z$};
\draw (1.5,-0.5) -- (2.25,-0.5);
\end{tikzpicture} 
 =
 \begin{tikzpicture}[baseline = (X.base),every node/.style={scale=0.6},scale=.4]
\draw[rounded corners] (3, 4) rectangle (4.2,-1);
\draw (3.6,1.5) node {$L_Z^\site{W}$};
\draw (4.2,3.5)  -- (6.7,3.5);
\draw (4.2,1.5)  -- (6.7,1.5);
\draw (4.2,-0.5) -- (5.2,-0.5);
edge\draw (5.2,-0.5) edge[out=0,in=0] (5.2,-1.5); 
\draw (6.2,-.5) -- (6.7,-.5); 
\draw (6.2,-0.5) edge[out=180,in=180] (6.2,-1.5); 
\draw[rounded corners] (6.7,4) rectangle (7.7,3);
\draw (7.2,3.5) node {$Z_C$};
\draw (7.2,3.0) -- (7.2,2.0); 
\draw[rounded corners] (6.7,2.0) rectangle (7.7,1);
\draw (7.2,1.5) node(X) {$W_Z$};
\draw (7.2,1.0) -- (7.2,-0.); 
\draw[rounded corners] (6.7,0.0) rectangle (7.7,-1);
\draw (7.2,-0.5) node {$\bar{V}_{Z_R}$};
\draw (7.7,1.5) -- (8.2,1.5); 
\draw (7.7,3.5) -- (8.2,3.5); 
\draw (7.7,-.5) -- (8.2,-.5); 
\draw (7.7,1.5) -- (8.7,1.5); 
\draw (7.7,3.5) -- (8.7,3.5); 
\draw (7.7,-.5) -- (8.7,-.5); 
\draw[rounded corners] (8.7, 4) rectangle (9.9,-1);
\draw (9.3,1.5) node {$R_{Z}^\site{W}$};
\end{tikzpicture}. 
\label{eq:X_Z_final}
\end{equation}
\begin{equation}
\begin{tikzpicture}[baseline = (X.base),every node/.style={scale=0.60},scale=.4]
\draw (-0.25,-0.5) -- (0.5,-0.5);
\draw[rounded corners] (0.5,0) rectangle (1.5,-1);  
\draw (1,-0.5) node (X) {$X_{B^i}$};
\draw (1.5,-0.5) -- (2.25,-0.5);
\end{tikzpicture} 
=
 \begin{tikzpicture}[baseline = (X.base),every node/.style={scale=0.6},scale=.4]
\draw (5,1.5) node (X) {$\phantom{X}$};,
\draw[rounded corners] (3, 4) rectangle (4.2,-1);
\draw (3.6,1.5) node {$L_A^\site{W}$};
\draw (4.2,3.5)  -- (5.2,3.5);
\draw (4.2,1.5)  -- (5.2,1.5);
\draw (4.2,-0.5) -- (5.2,-0.5);
\draw[rounded corners] (5.2, 4) rectangle (7.2,-1);
\draw (6.25,1.5) node {$\displaystyle\prod_{j=1}^{i-1} E_{B^j_L}^\site{W}$};
\draw (7.2,3.5) -- (8.2,3.5);
\draw (7.2,1.5) -- (8.2,1.5);
\draw (7.2,-.5) -- (8.2,-0.5);
\draw[rounded corners] (8.2,4) rectangle (9.2,3);
\draw (8.7,3.5) node {$B^i_C$};
\draw (8.7,3.0) -- (8.7,2.0); 
\draw[rounded corners] (8.2,2.0) rectangle (9.2,1);
\draw (8.7,1.5) node {$W_i$};
\draw (8.7,1.0) -- (8.7,-0.); 
\draw[rounded corners] (8.2,0.0) rectangle (9.2,-1);
\draw (8.7,-0.5) node {$\bar{V}_{B^i_L}$};
\draw (9.2,1.5) -- (11.7,1.5); 
\draw (9.2,3.5) -- (11.7,3.5); 
\draw (9.2,-.5) -- (9.7,-.5); 
edge\draw (9.7,-0.5) edge[out=0,in=0] (9.7,-1.5); 
\draw (10.7,-0.5) edge[out=180,in=180] (10.7,-1.5); 
\draw (10.7,-.5) -- (11.7,-.5); 
\draw[rounded corners] (11.7, 4) rectangle (14.3,-1);
\draw (13.05,1.5) node {$\displaystyle\prod_{j=i+1}^n E_{B^j_R}^\site{W}$};
\draw (14.3,3.5) -- (15.3,3.5);
\draw (14.3,1.5) -- (15.3,1.5);
\draw (14.3,-.5) -- (15.3,-0.5);
\draw (5,1.5) node (X) {$\phantom{X}$};,
\draw[rounded corners] (15.3, 4) rectangle (16.5,-1);
\draw (15.9,1.5) node {$R_Z^\site{W}$};
\end{tikzpicture}, 
\label{eq:X_B_final}
\end{equation}
for $i = 1, \cdots, n-1$. 
\begin{equation}
\begin{tikzpicture}[baseline = (X.base),every node/.style={scale=0.60},scale=.4]
\draw (0,-0.5) -- (0.5,-0.5);
\draw[rounded corners] (0.5,0) rectangle (1.5,-1);  
\draw (1,-0.5) node {$b_L^n$};
\draw (1,-0.8) node(X) {};
\draw (1,-1) -- (1,-1.5);
\draw (1.5,-0.5) -- (2,-0.5);
\end{tikzpicture} 
=  
\begin{tikzpicture}[baseline = (X.base),every node/.style={scale=0.6},scale=.4]
\draw (5,1.5) node (X) {$\phantom{X}$};,
\draw[rounded corners] (3, 4) rectangle (4.2,-1);
\draw (3.6,1.5) node {$L_A^\site{W}$};
\draw (4.2,3.5)  -- (5.2,3.5);
\draw (4.2,1.5)  -- (5.2,1.5);
\draw (4.2,-0.5) -- (5.2,-0.5);
\draw[rounded corners] (5.2, 4) rectangle (7.2,-1);
\draw (6.25,1.5) node {$\displaystyle\prod_{i=1}^{n-1} E_{B^i_L}^\site{W}$};
\draw (7.2,3.5) -- (8.2,3.5);
\draw (7.2,1.5) -- (8.2,1.5);
\draw (7.2,-.5) -- (7.7,-0.5);
edge\draw (7.7,-0.5) edge[out=0,in=0] (7.7,-1.5); 
\draw[rounded corners] (8.2,4) rectangle (9.2,3);
\draw (8.7,3.5) node {$B^n_C$};
\draw (8.7,3.0) -- (8.7,2.0); 
\draw[rounded corners] (8.2,2.0) rectangle (9.2,1);
\draw (8.7,1.5) node {$W_n$};
\draw (8.7,1.0) -- (8.7,-1.5); 
\draw (9.2,3.5) -- (10.2,3.5);
\draw (9.2,1.5) -- (10.2,1.5);
\draw (9.7,-.5) -- (10.2,-0.5);
\draw (9.7,-0.5) edge[out=180,in=180] (9.7,-1.5); 
\draw[rounded corners] (10.2, 4) rectangle (11.4,-1);
\draw (10.8,1.5) node {$R_Z^\site{W}$};
\end{tikzpicture}. 
\label{eq:bn_final}
\end{equation} 
We can now put Eq. \ref{eq:X_A_final}-\ref{eq:bn_final} back into Eq. \ref{eq:Phi} to obtain $\ket{\Phi}_{H} = \text{Proj}_{T\PH} \HPsi$.    

$X_A$ contains no information about $B^i$ and $Z$, and in fact, is exactly the same effective parameter as in a translationally invariant system composed of only $A$ and $W_A$ \cite{Tangent_space, iTDVP}.     
Thus, the bulk tensors $A$ and $Z$ should evolve as if they are in an entirely uniform MPS, by the iTDVP algorithm in \cite{Tangent_space, iTDVP}.    
The effect of the left and the right bulks on the $B$ tensors only comes through the boundary tensors $\rbra{L_A^\site{W}}$ and $\rket{R_Z^\site{W}}$. 
In fact, in a finite system parametrized only by the $B$ tensors, the tensors at the left (right) boundary have no left (right) indices, and the effective parameters are given by the terms in Eq. \ref{eq:X_B_final} and \ref{eq:bn_final} without the $\rbra{L_A^\site{W}}$ and $\rket{R_Z^\site{W}}$ tensors \cite{finite_TDVP}.   
Thus, the $B$ matrices can be evolved by the same TDVP algorithm in \cite{finite_TDVP} of a finite system, except under the additional influence of $\rbra{L_A^\site{W}}$ and $\rket{R_Z^\site{W}}$.    
The only thing unclear is how to patch the time evolutions of $A, B^i$ and $Z$ together, which we explain in the next section. 
\section{Integration scheme}
\label{sec:integration}
\begin{table}
\caption{Pseudocode of mixed-iTDVP for step $\delta t$.}
\begin{minipage}{1.0\linewidth}
\begin{algorithm}[H]
  \caption{Mixed-iTDVP: evolving $\ket{\Psi}$ to $e^{\delta t \hat{H}}\ket{\Psi}$}
\label{alg:Heff_NN}
\begin{algorithmic}[1]
\Require MPO tensor $W_A$, $W_1, \cdots, W_{n_W}$, $W_Z$; MPS tensor $\{A_L, A_R, C_A, A_C\}$, $\{Z_L, Z_R, C_Z, Z_C\}$, $B^1_C, B^2_R, \cdots, B^n_R$; $L_A^\site{W}$, $R_Z^\site{W}$; time step $\delta t$  
  \Ensure MPS tensor $\{A_L, A_R, C_A, A_C\}$, $\{Z_L, Z_R, C_Z, Z_C\}$, $B^1_C, B^2_R,\cdots,B^n_R$; $L_A^\site{W}$, $R_Z^\site{W}$
  \State \{$A_L,A_R,C_A,A_C$\} $\gets$ iTDVP($W_A$,$A_L,A_R,C_A,A_C,\delta t$)
  \State Compute $L_A^\site{W}$ with $A_L$ and $W_A$  
  \State \{$B^1_L,\cdots,B^{n-1}_L,B^{n}_C$\} $\gets$ right sweep of finite-size TDVP($B^1_C,B^2_R,\cdots,B^{n}_R$,$L_A^\site{W}$,$R_Z^\site{W}$,$\delta t/2$) 
  \State \{$Z_L,Z_R,C_Z,Z_C$\} $\gets$ iTDVP($W_Z$,$Z_L,Z_R,C_Z,Z_C,\delta t$)
  \State Compute $R_Z^\site{W}$ with $Z_R$ and $W_Z$ 
  \State \{$B^1_C,B^2_R,\cdots,B^{n}_R$\} $\gets$ left sweep of finite-size TDVP($W_1,\cdots,W_n$,$B^1_L,\cdots,B^{n-1}_L,B^{n}_C$,$L_A^\site{W}$,$R_Z^\site{W}$,$\delta t/2$) 
\end{algorithmic}
\end{algorithm}
\end{minipage}
\label{tab:mixed-iTDVP}
\end{table}
Here we explain how to evolve $\ket{\Psi}$ to $e^{\delta t \hat{H}} \ket{\Psi}$ using $\ket{\Phi}_H$. 
In iTDVP, one first puts the center site $A_C$ at left infinity.  
Then one exponentiates the terms in $\ket{\Phi}_H$, one by one from left to right, to sequentially act on the current state.  
As the algorithm sweeps from left infinity to site $0$, the effect of the left boundary tensor decays away and the $A_C$ and $C_A$ tensors converge to their respective limits.  
The iTDVP algorithm in \cite{iTDVP} finds these limits without doing the actual sweep, and is thus very efficient. 
However, there is something very peculiar about the sweeping process:  in obtaining $\{A_C(t+\delta t), C_A(t+\delta(t))\}$ from $\{A_C(t), C_A(t)\}$, when the action of one term in $\ket{\Phi}_H$ is completed, one ends up with $C_A(t)$ instead of $C_A(t+\delta t)$ as the bond matrix.   
(One step of the sweep consists of two half-steps, and $C_A(t+\delta t)$ is obtained after the first half-step.) 
See page 35 of \cite{Tangent_space} or Table 1 of \cite{iTDVP} for the details. 
This peculiar fact is the key to patch the iTDVP and the finite TDVP algorithms. 

Suppose that at time $t$, we have a mixed iMPS centered at $B_C^1(t)$: 
\begin{equation*}
 \dots
\begin{tikzpicture}[baseline = (X.base),every node/.style={scale=0.6},scale=.4]
\draw (0.5,1.5) -- (1,1.5); 
\draw[rounded corners] (1,2) rectangle (3.4,1);
\draw (2.2,1.5) node (X) {$A_L(t)$};
\draw (2.2,1) -- (2.2,.5); 
\draw (3.4,1.5) -- (10.0,1.5); 
\draw[rounded corners] (10.0,2) rectangle (11.8,1);
\draw (10.9,1.5) node {$B_C^1(t)$};
\draw (10.9,1.0) -- (10.9,0.5); 
\draw (11.8,1.5) -- (12.3,1.5);
\draw (12.8,1.5) node[scale=1.25] (X) {\dots};
\draw (13.3,1.5) -- (13.8,1.5);
\draw[rounded corners] (13.8,2) rectangle (15.6,1);
\draw (14.7,1.5) node {$B_R^{n}(t)$};
\draw (14.7,1.0) -- (14.7,0.5); 
\draw (15.6,1.5) -- (16.6,1.5);
\draw[rounded corners] (16.6,2) rectangle (18.4,1);
\draw (17.5,1.5) node {$Z_R(t)$};
\draw (17.5,1.0) -- (17.5,0.5); 
\draw (18.4,1.5) -- (18.9,1.5);
\end{tikzpicture} \dots
\end{equation*}
To make the MPS centered at $A_C(t)$ at left infinity, one needs to borrow a $C_A(t)$ from $B^1_C(t)$, so that one has
\begin{equation*}
 \dots
\begin{tikzpicture}[baseline = (X.base),every node/.style={scale=0.6},scale=.4]
\draw (0.5,1.5) -- (1,1.5); 
\draw[rounded corners] (1,2) rectangle (3.4,1);
\draw (2.2,1.5) node (X) {$A_R(t)$};
\draw (2.2,1) -- (2.2,.5); 
\draw (3.4,1.5) -- (7.2,1.5); 
\draw[rounded corners] (7.2,2) rectangle (9.0,1);
\draw (8.1,1.5) node {$C_A^{-1}(t)$};
\draw (9.0,1.5) -- (10.0,1.5);
\draw[rounded corners] (10.0,2) rectangle (11.8,1);
\draw (10.9,1.5) node {$B_C^1(t)$};
\draw (10.9,1.0) -- (10.9,0.5); 
\draw (11.8,1.5) -- (12.3,1.5);
\draw (12.8,1.5) node[scale=1.25] (X) {\dots};
\draw (13.3,1.5) -- (13.8,1.5);
\draw[rounded corners] (13.8,2) rectangle (15.6,1);
\draw (14.7,1.5) node {$B_R^{n}(t)$};
\draw (14.7,1.0) -- (14.7,0.5); 
\draw (15.6,1.5) -- (16.6,1.5);
\draw[rounded corners] (16.6,2) rectangle (18.4,1);
\draw (17.5,1.5) node {$Z_R(t)$};
\draw (17.5,1.0) -- (17.5,0.5); 
\draw (18.4,1.5) -- (18.9,1.5);
\end{tikzpicture} \dots
\end{equation*}
One then performs iTDVP on $A$ for $\delta t$ to arrive at 
\begin{equation*}
 \dots
\begin{tikzpicture}[baseline = (X.base),every node/.style={scale=0.6},scale=.4]
\draw (0.5,1.5) -- (1,1.5); 
\draw[rounded corners] (1,2) rectangle (3.4,1);
\draw (2.2,1.5) node (X) {$A_L(t+\delta t)$};
\draw (2.2,1) -- (2.2,.5); 
\draw (3.4,1.5) -- (4.4,1.5); 
\draw[rounded corners] (4.4,2) rectangle (6.2,1);
\draw (5.3,1.5) node {$C_A(t)$};
\draw (6.2,1.5) -- (7.2,1.5);
\draw[rounded corners] (7.2,2) rectangle (9.0,1);
\draw (8.1,1.5) node {$C_A^{-1}(t)$};
\draw (9.0,1.5) -- (10.0,1.5);
\draw[rounded corners] (10.0,2) rectangle (11.8,1);
\draw (10.9,1.5) node {$B_C^1(t)$};
\draw (10.9,1.0) -- (10.9,0.5); 
\draw (11.8,1.5) -- (12.3,1.5);
\draw (12.8,1.5) node[scale=1.25] (X) {\dots};
\draw (13.3,1.5) -- (13.8,1.5);
\draw[rounded corners] (13.8,2) rectangle (15.6,1);
\draw (14.7,1.5) node {$B_R^{n}(t)$};
\draw (14.7,1.0) -- (14.7,0.5); 
\draw (15.6,1.5) -- (16.6,1.5);
\draw[rounded corners] (16.6,2) rectangle (18.4,1);
\draw (17.5,1.5) node {$Z_R(t)$};
\draw (17.5,1.0) -- (17.5,0.5); 
\draw (18.4,1.5) -- (18.9,1.5);
\end{tikzpicture} \dots
\end{equation*}
Thus, the bond matrix $C_A(t)$ cancels, and one next carries out the right sweep of the finite TDVP algorithm on $B$ for $\delta t/2$ with boundary tensors $\rbra{L_{A(t+\delta t)}^\site{W}}$ and $\rket{R_{Z(t)}^\site{W}}$. 
Then one does iTDVP on $Z$ for $\delta t$ and sweeps on $B$ leftward for $\delta t/2$ with boundary tensors $\rbra{L_{A(t+\delta t)}^\site{W}}$ and $\rket{R_{Z(t+\delta t)}^\site{W}}$. 
This completes the mixed-iTDVP for one step of $\delta t$. 
For a pseudocode, see Table \ref{tab:mixed-iTDVP}. 
We call this algorithm {\it mixed-iTDVP}. 
Globally, mixed-iTDVP is second order in $\delta t$ if $A$ and $Z$ are eigenstates of the bulk Hamiltonian on the left and right, which is the same as the finite TDVP algorithm. 
It is first order in $\delta t$ if $A$ and $Z$ evolve non-trivially, which results from the iTDVP algorithm. 

The algorithm can also be used to find the ground state when $\delta t$ is real and negative. 
When the time step is infinite, the algorithm reduces to the conventional one-site density matrix renormalization group \cite{DMRG}.
When the time step approaches 0, however, the time-evolution algorithm has the benefit of ensuring finding the global energy minimum, as long as the initial state has non-zero overlap with the ground state.   

To dynamically expand $n$, simply upgrade some number of $A$ and $Z$ matrices to be part of $B$. 
The procedure used in Sec. \ref{sec:Ising} is that, during the time-evolution process, when the half-chain entanglement entropy at $B^{i=5}$ differs from that at $A$ by more than $10^{-5}$, we add five more $B$ tensors equal to $A$ to the left end of the inhomogeneous region. 
The same is done to the right, too.   
The fact that $n$ can be expanded dynamically means that one can start with a very small inhomogeneous region at the early times of the time evolution and expand it gradually as time increases.  
This is an advantage compared to a finite-size algorithm.  
\section{Example: quantum Ising model}
\label{sec:Ising}
As an illustrative example, we study the quantum dynamics of the quantum Ising chain: 
\begin{equation}
  \hat{H}_\Ising = J\sum_{i=-\infty}^\infty \hat{\sigma}^z_i \hat{\sigma}^z_{i+1} + \sum_{i=-\infty}^{\infty} (h_x \hat{\sigma}^x_i + h_z \hat \sigma^z_i) 
  \label{eq:Ising}
\end{equation}
where $\hat\sigma^{x,z}$ are the Pauli matrices. 
It is integrable when $h_z = 0$ or $h_x = 0$, and is critical when $h_z$ = 0 and $h_x/J = \pm1$ \cite{Kogut}.
At criticality, the dispersion relation becomes linear: $E(\vec k) = v_s \abs{\vec k}$, giving a characteristic sound velocity $v_s = 2$ \cite{Kogut}.
We denote the pre-quenched Hamiltonian by $\hat H_0$ and the post-quenched Hamiltonian by $\hat H_1$. 
In the following, $\hat H_1 = \hat H_0 + \delta \hat H$, where $\delta \hat H$ is a local field on site $i_0$ at the middle of region $B$. 
When the quench is local, we observe that the entanglement entropy saturates at long time.  
This means that one can study the quantum dynamics for long times with a relatively small bond dimension, well into the stationary limit. 
\subsection{Benchmark}
\begin{figure}[htb]
\centering
\caption{ 
$\braket{\hat\sigma^x_{i_0}(t)}$ as a function of time, $\hat H_0$ with $J$ = -1, $h_x = 1.5$, and $h_z = 0$, and $\delta \hat {H} = \hat\sigma^x_{i_0}$. 
The mixed-iTDVP computation is done with $\delta t = 0.005$ and $D = 20$.
The exact Ising solution is computed for an open chain with 512 sites.  
The inset is $\braket{\hat\sigma^x_{i_0}(t)}$ from $t = 40$ to 50. 
}
\includegraphics[scale=0.27]{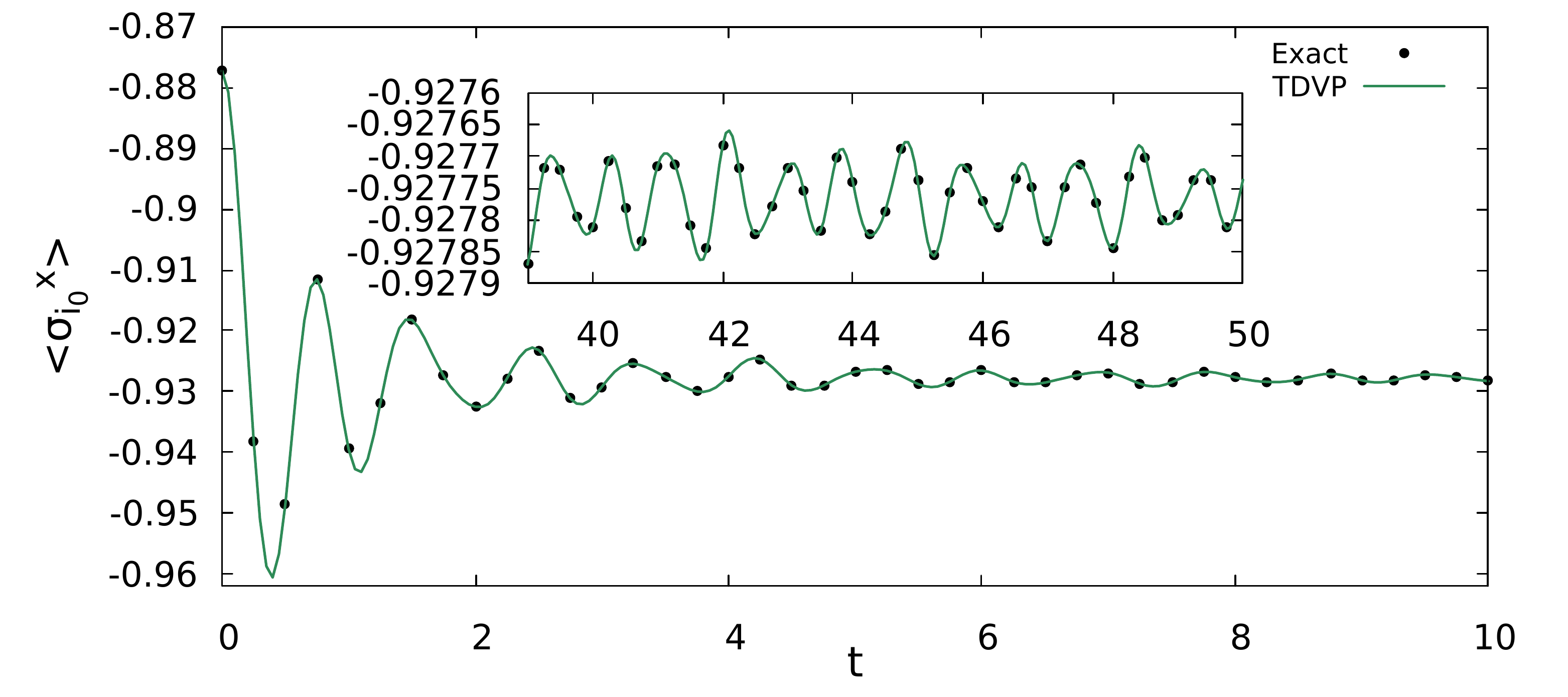}
\label{fig:X}
\end{figure}

We benchmark our algorithm with $\hat H_0$ with $J$ = -1, $h_x = 1.5$, and $h_z = 0$, and $\delta \hat {H} = \hat\sigma^x_{i_0}$. 
This local quench does not break the integrability of the transverse-field Ising chain, and thus the quench dynamics can be computed exactly on a finite chain.  
We follow \cite{Ising_solution} to compute the quench dynamics. 
In Fig. \ref{fig:X}, we show the transverse magnetization at site $i_0$, $\braket{\hat\sigma^x_{i_0}(t)}$, as a function of time, obtained both with mixed-iTDVP and the Ising exact solution. 
As seen, the mixed-iTDVP works correctly well into the stationary regime. 
\subsection{Effect of finite size}
\begin{figure}[htb]
\centering
\caption{ 
$\braket{\hat\sigma^x_i}$ in the ground state of $\hat H_{\Ising}$ + $\hat\sigma_{i_0}^z$ with $J = -1, h_x = 1.05$, and $h_z = 0$. 
The calculation is done with $D = 20$. 
The inset is a zoomed-in version of the main plot. 
The curves for the finite system and the infinite system are overlapping for most times.  
}
\includegraphics[scale=0.27]{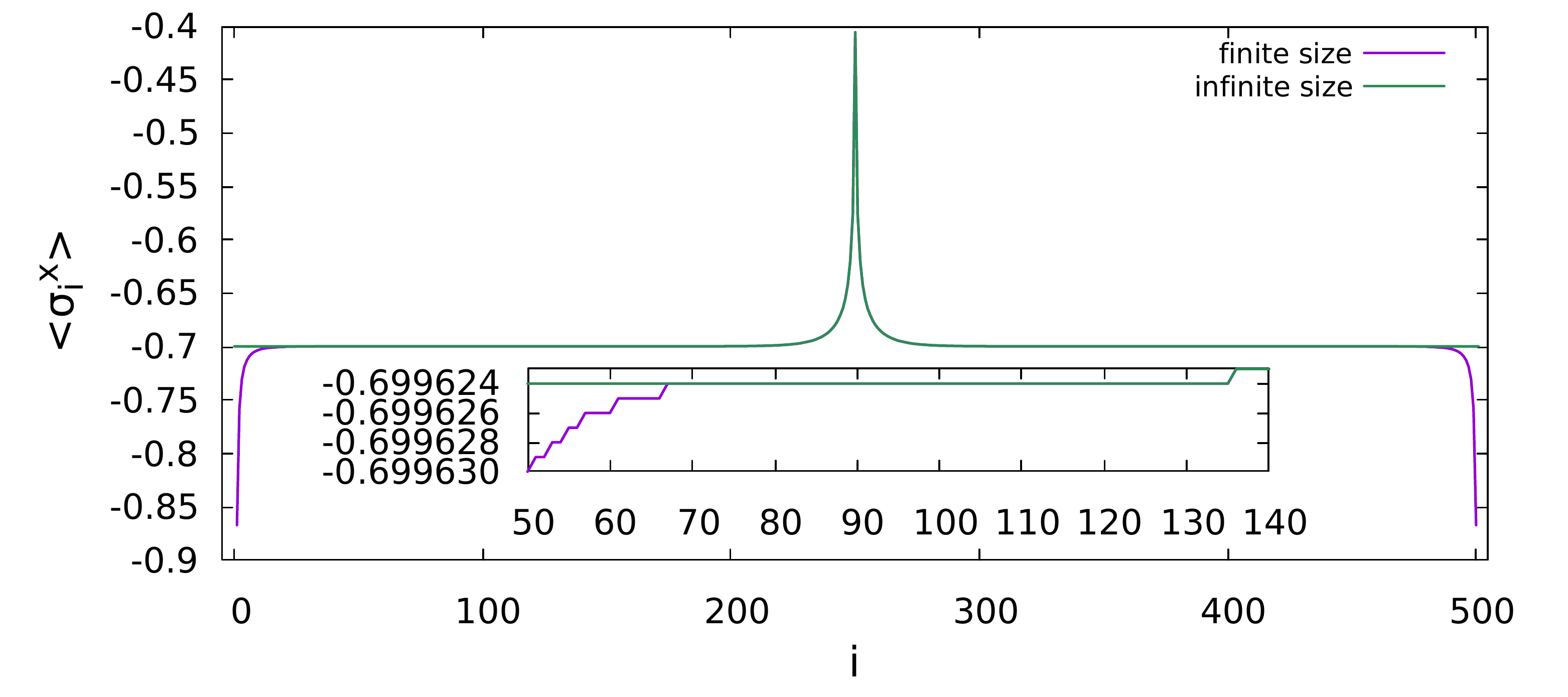}
\label{fig:size}
\end{figure}

The defining feature of mixed-iTDVP is that it works directly in the thermodynamic limit. 
We demonstrate the lack of the finite size effect by computing the ground state of an inhomogeneous Hamiltonian: $\hat H_{\Ising}$ + $\hat\sigma_{i_0}^z$ with $J = -1, h_x = 1.05$, and $h_z = 0$, where $i_0$ is in the middle of the chain. 
The transverse magnetization of the ground state is shown in Fig. \ref{fig:size}, in comparison with a finite size calculation with 500 sites. 

\subsection{speed of information spreading}
Here we consider the spread of information after a local quench in the Ising chain both in the ballistic and the diffusive case.  
In the ballistic case, the system is integrable and admits an extensive number of non-interacting quasi-particles in its spectrum, which transports energy ballistically. 
When both $h_x$ and $h_z$ are non-zero, however, the Ising chain is no longer integrable, and the only locally conserved quantity is the energy.   
In this case, there are no ballistically propagating quasi-particles so that, in an extended quantum quench, the energy is transported in a way similar to a random walk, at a speed which is proportional to $\sqrt{t}$ \cite{Ballistic_EE}.  
This is called a diffusive system. 

\begin{figure}[htb]
\centering
\caption{ 
$\braket{\hat\sigma^x_i}$ as a function of time, represented in a curve plot for both the ballistic system (top): $\hat H_0$ = $\hat H_\Ising$ with $J = -1, h_x = 1.5$, and $h_z = 0$, and the diffusive system (bottom): $\hat H_0$ = $\hat H_\Ising$ with $J = 1$, $h_x$ = 0.9045, and $h_z$ = 0.8090. 
The quenching Hamiltonian is $\delta \hat H = \hat \sigma^z_{i_0}$ in both cases. 
The computation is done with $\delta t = 0.005$ and $D = 20$. 
Computations with $D = 30$ are also done, and the results are well-converged with the bond dimension. 
}
\begin{subfigure}{0.5\textwidth}
  \caption{Ballistic system}
\includegraphics[scale=0.27]{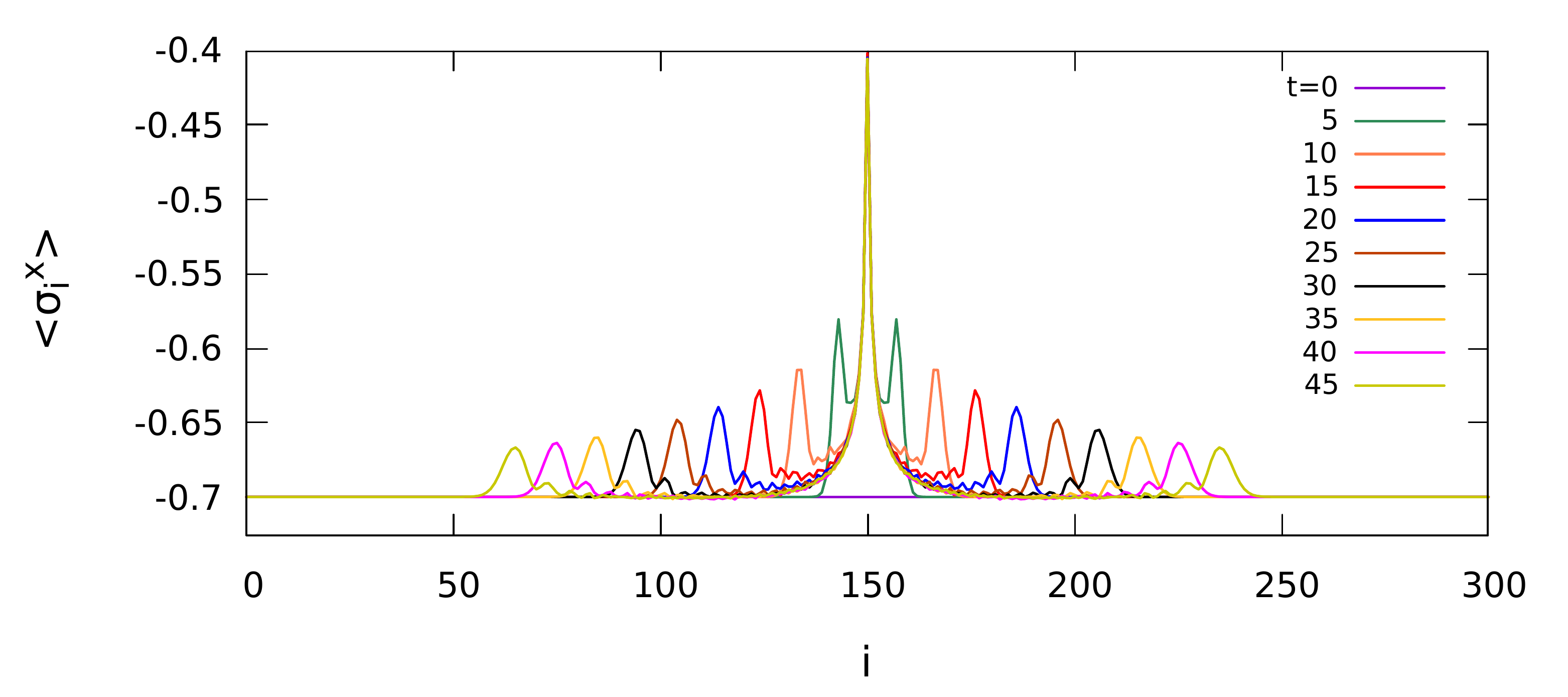}
\end{subfigure}
\begin{subfigure}{0.5\textwidth}
  \caption{Diffusive system}
\includegraphics[scale=0.27]{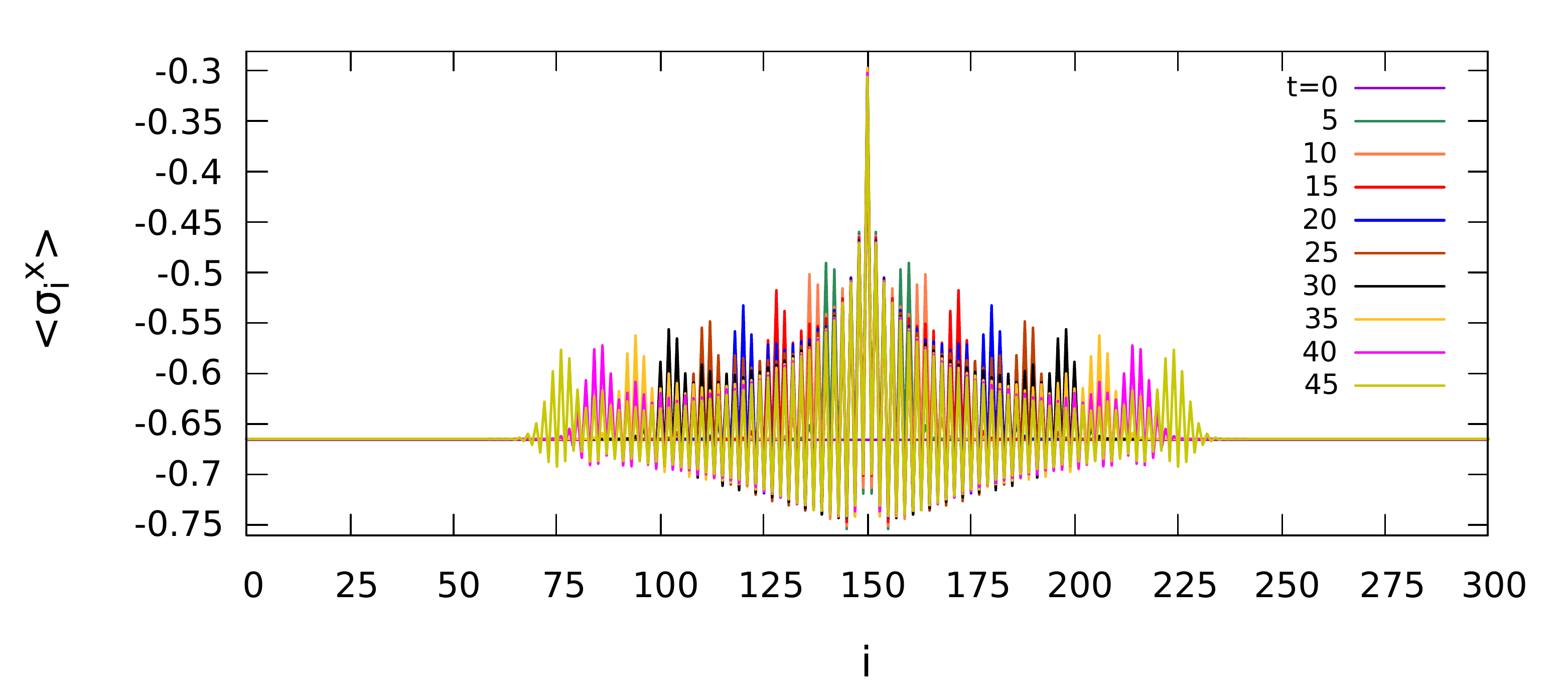}
\end{subfigure}
\label{fig:curve}
\end{figure}

\begin{figure}[htb]
\centering
\caption{ 
$\braket{\hat\sigma^x_i}$ as a function of time, represented in a contour plot, for the same two quenches described in Fig. \ref{fig:curve}. 
}
\begin{minipage}{.24\textwidth}
  \caption*{Ballistic system}
  \includegraphics[scale=0.37]{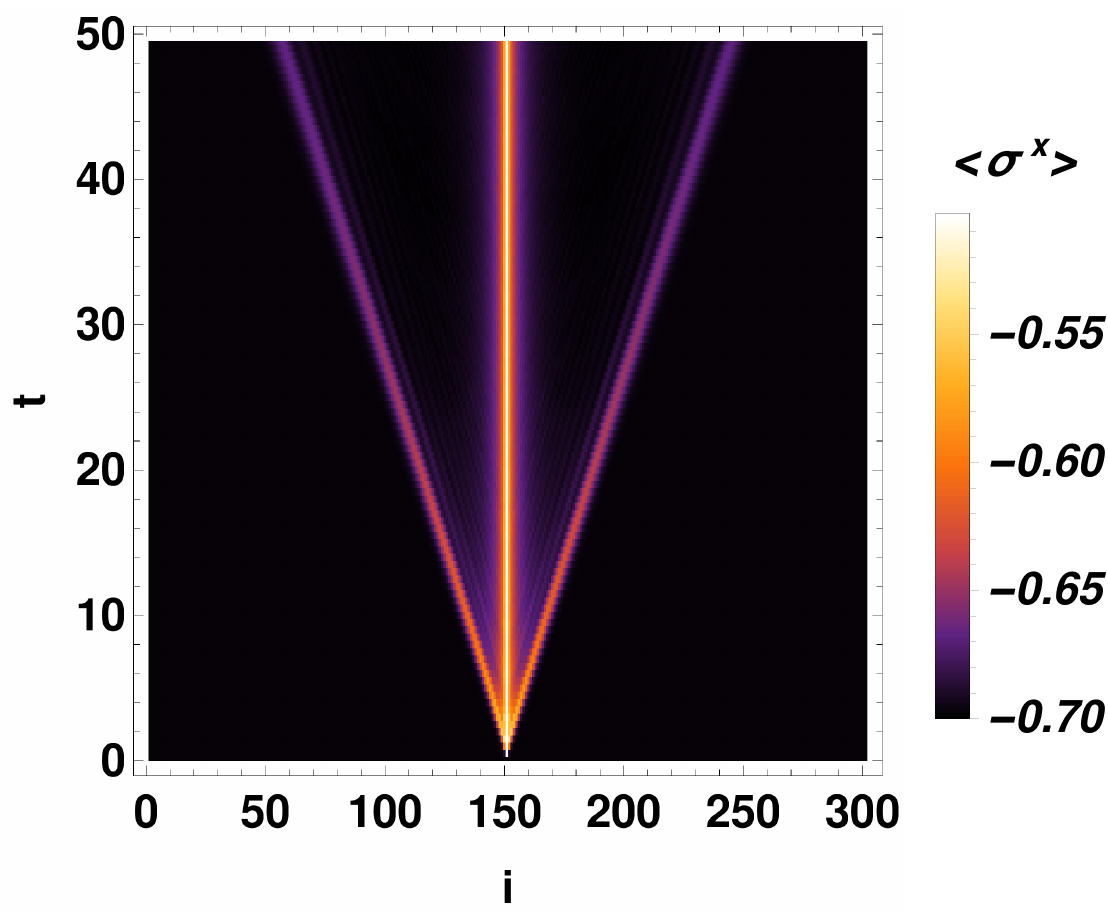}
\end{minipage}%
\begin{minipage}{.24\textwidth}
  \caption*{Diffusive system}
  \includegraphics[scale=0.37]{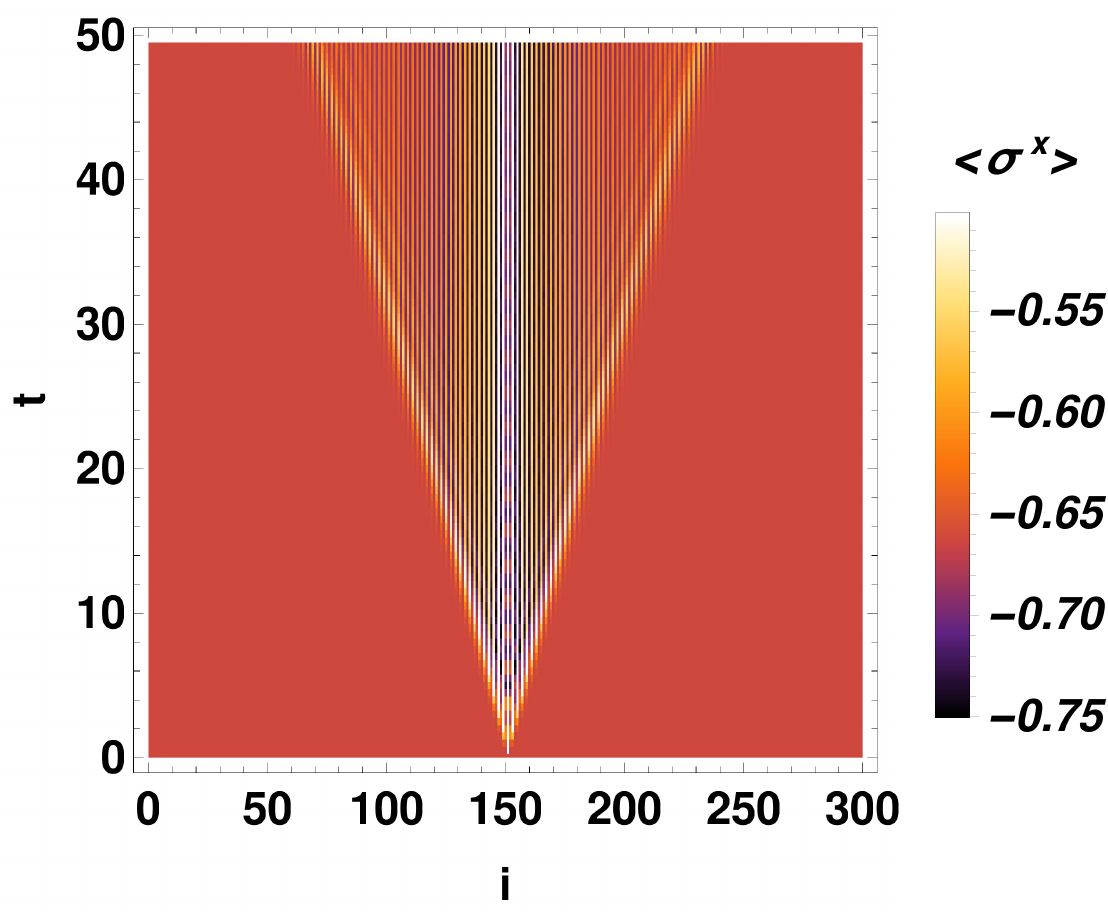}
\end{minipage}%
\label{fig:contour}
\end{figure}

For the ballistic case, we take $\hat H_0$ to be the $\hat H_\Ising$ with $J = -1$, $h_x = 1.5$, and $h_z = 0$.   
For the diffusive case, we take $\hat H_0$ to be the $\hat H_\Ising$ with $J = 1$, $h_x = 0.9045$, and $h_z = 0.8090$, which is shown to be robustly non-integrable in \cite{Ballistic_EE}.   
In both cases, the local quench is done through $\delta \hat H = \hat\sigma^z_{i_0}$, where we place $i_0$ in the middle of the inhomogeneous region $B$.  
To monitor the spread of information, we measure the time dependence of $\braket{\hat\sigma^x_i}$ on the whole chain, shown in Fig. \ref{fig:curve} and \ref{fig:contour}. 
The time dependence of other local observables are similar with $\braket{\hat \sigma_i^x}$. 

A very sharp wave-front is observed in both cases as the information of the local quench spreads. 
While this is expected for the ballistic system, it is surprising for the diffusive system, because the energy transports only diffusively in an extended quench. 
The slope of the wave-front can be computed to give the speed of information spreading, $v_w$.  
More specifically, we do a linear fit of the function $i_\text{ridge}(t)$, which for the ballistic system equals the site of the left-most local maximum of $\braket{\hat \sigma^x_i}$ at time $t$, and take the slope of the linear fit as the slope of the wave-front.  
For the diffusive system, we note that there exists a secondary peak in the magnetization profile, for example at around $i = 75$ at $t$ = 45. 
We take the $i_\text{ridge}(t)$ to be the site on which $\braket{\hat\sigma_i^x}$ is the largest in this secondary peak.  

The fitted $v_w$s are shown in Table \ref{tab:vw}.  
\begin{table}
  \setlength{\tabcolsep}{1.2em}
\centering
  \caption{Velocity of the wave-front in local quenches.
  The number in the parenthesis is the uncertainty of the fit on the last digit. 
  $R^2$ is the $R$-square of the linear fit.}
\begin{tabular}{lllll} 
  \hline
  \hline
  system     & $D$ & $v_w$     & $R^2$\\
  \hline
  ballistic  & 20  & 1.94(2)   &0.99979\\ 
  diffusive  & 20  & 1.71(3)   &0.99949\\ 
  \hline
  \hline
\end{tabular}
\label{tab:vw}
\end{table}
Because of the discrete nature of $i$, $i_\text{ridge}(t)$ can be ambiguous up to $\pm 1$. 
This contributes to the slight non-linearity of $r_\text{ridge}(t)$, indicated by $R^2 < 1$. 
For the ballistic system, there are well-defined quasi-particles whose velocities are given by the dispersion relation: $E(k) = 2\sqrt{1-2h_x\cos(k)+h_x^2}$ \cite{Kogut}.   
One thus expects that the speed of information spreading should be 
\begin{equation}
  v_w = \max_{k\in[-\pi,\pi]} \frac{dE(k)}{dk}
\end{equation}
which equals 2 for all $h_x$ for the transverse-field Ising model. 
This is very close to the velocity actually measured in the local quench. 
The presence of the light-cone in the diffusive system, however, suggests that the ballistic spread of information is generic in a local quench, and happens not only in integrable systems.   
\section{Discussion}
\label{sec:discussion}
In this paper, we gave a detailed derivation of the TDVP equation for mixed infinite MPSs.  
The result is a simple combination of the finite TDVP and infinite TDVP algorithms, both of which are inversion-free. 
The method was applied to local quenches of the quantum Ising model, and interesting phenomenon were found, which calls for future work. 
We also expect future work on the algorithmic side. 
For example, we note that the mixed infinite MPS is very similar to the variational ansatz of the elementary excitations \cite{Tangent_space} of a translationally invariant system: 
\begin{equation}
  \ket{\Psi_k} =\sum_{x} e^{ikx} \dots
\begin{tikzpicture}[baseline = (X.base),every node/.style={scale=0.6},scale=.4]
\draw (0.5,1.5) -- (1,1.5); 
\draw[rounded corners] (1,2) rectangle (2,1);
\draw (1.5,1.5) node (X) {$A$};
\draw (2,1.5) -- (3,1.5); 
\draw[rounded corners] (3,2) rectangle (4,1);
\draw (3.5,1.5) node {$A$};
\draw (4,1.5) -- (5,1.5);
\draw[rounded corners] (5,2) rectangle (6,1);
\draw (5.5,1.5) node {$B^x$};
\draw (5.5,0.25) node {$x$};
\draw (1.5,1) -- (1.5,.5); 
\draw (3.5,1) -- (3.5,.5); 
\draw (5.5,1) -- (5.5,.5);
\draw (6,1.5) -- (7.0,1.5); 
\draw[rounded corners] (7,2) rectangle (8,1);
\draw (7.5,1.5) node (X) {$Z$};
\draw (8,1.5) -- (9,1.5); 
\draw[rounded corners] (9,2) rectangle (10,1);
\draw (9.5,1.5) node (X) {$Z$};
\draw (10,1.5) -- (10.5,1.5); 
\draw (7.5,1) -- (7.5,.5); 
\draw (9.5,1) -- (9.5,.5);
\end{tikzpicture} \dots, 
\label{eq:momentum}
\end{equation}
where $x$ labels the position of spin sites. 
We thus hope that the current method can help develop a time-evolution algorithm for the elementary excitations. 

\begin{acknowledgments}
  The code is based on ITensor \cite{ITensor} (version 3, C\texttt{++}), and is available upon request. 
  The author is grateful for the help received on the ITensor Support Q\&A. 
He is grateful for mentorship from his advisor Roberto Car at Princeton. 
He acknowledges support from the DOE Award DE-SC0017865. 
\end{acknowledgments}

\section{Appendix}
\subsection{Derivation of Eq. \ref{eq:X_A_final}-\ref{eq:bn_final}}
\label{app:ortho}
Before we start, we need some facts about the MPS transfer matrices $\EL$ and $\ER$, defined as 
\begin{equation}
  \EL = 
  \begin{tikzpicture}[baseline = (X.base),every node/.style={scale=0.6},scale=.4]
\draw (0.5, 1.5) -- (1,1.5);
\draw (0.5, -1.5) -- (1,-1.5);
\draw[rounded corners] (1,2) rectangle (2,1);
\draw[rounded corners] (1,-1) rectangle (2,-2);
\draw (1.5,1) -- (1.5,-1);
\draw (1.5,1.5) node {$A_L$};
\draw (1.5,0) node(X) {};
\draw (1.5,-1.5) node {$\bar{A}_L$};
\draw (2,1.5) -- (2.5,1.5); 
\draw (2,-1.5) -- (2.5,-1.5);
\end{tikzpicture} 
\hspace{10mm}
  \ER = 
  \begin{tikzpicture}[baseline = (X.base),every node/.style={scale=0.6},scale=.4]
\draw (0.5, 1.5) -- (1,1.5);
\draw (0.5, -1.5) -- (1,-1.5);
\draw[rounded corners] (1,2) rectangle (2,1);
\draw[rounded corners] (1,-1) rectangle (2,-2);
\draw (1.5,1) -- (1.5,-1);
\draw (1.5,1.5) node {$A_R$};
\draw (1.5,-1.5) node {$\bar{A}_R$};
\draw (2,1.5) -- (2.5,1.5); 
\draw (2,-1.5) -- (2.5,-1.5);
\end{tikzpicture}. 
\label{eq:E}
\end{equation}
We note that the canonical condition, Eq. \ref{eq:canonical}, is the eigen-relation for the non-degenerate leading eigenvalue of the transfer operators, which is 1 for a normalized uniform MPS \cite{Tangent_space}.  
This is very important, because it means that if one propagates an arbitrary boundary tensor from left through infinitely many $\EL$, only the leading left-eigvector of $\EL$ survives, which is a two-index delta tensor. 
The analogous fact is true for $\ER$, too. 

We now determine the $\ket{\Phi({X}_A;{X}_{B^i};{X}_Z;{b}^{n}_L)}$
that is the orthogonal projection of $\HPsi$ on the tangent space at $\ket{\Psi}$.     
To do this, we need to first compute the inner product $\braket{\Phi|\Phi}$, also known as the Gram matrix.
Using Eq. \ref{eq:canonical} and \ref{eq:Phi}-\ref{eq:V}, we have:  
\begin{equation*}
  \begin{split}
  \braket{\Phi(&\bar{X}_A;\bar{X}_{B^i};\bar{X}_Z;\bar{b}^{n}_L)|\Phi({X}_A;{X}_{B^i};{X}_Z;{b}_L^n)}
\\
  &= \sum_{m=0}^\infty
\begin{tikzpicture}[baseline = (X.base),every node/.style={scale=0.6},scale=.4]
\draw (1,-1.5) edge[out=180,in=180] (1,1.5);
\draw[rounded corners] (1,2) rectangle (2,1);
\draw[rounded corners] (1,-1) rectangle (2,-2);
\draw (1.5,1.5) node {$X_A$};
\draw (1.5,-1.5) node {$\bar{X}_A$};
\draw (2,1.5) -- (2.5,1.5); 
\draw (2,-1.5) -- (2.5,-1.5);
\draw[rounded corners] (2.5,2) rectangle (4.5,-2);
\draw (3.5,0) node {$(\ER)^m$};
\draw (4.5,1.5) -- (5.0,1.5);  
\draw (4.5,-1.5) -- (5.0,-1.5);
\draw[rounded corners] (5,2) rectangle (6,1);
\draw[rounded corners] (5,-1) rectangle (6,-2);
\draw (5.5,1) -- (5.5,-1);
\draw (5.5,1.5) node {$B^1_R$};
\draw (5.5,0) node(X) {};
\draw (5.5,-1.5) node {$\bar{B}^1_R$};
\draw (6,-1.5) edge[out=0,in=0] (6,1.5);
\end{tikzpicture} 
+ \sum_{m=0}^\infty
\begin{tikzpicture}[baseline = (X.base),every node/.style={scale=0.6},scale=.4]
\draw (1,-1.5) edge[out=180,in=180] (1,1.5);
\draw[rounded corners] (1,2) rectangle (2,1);
\draw[rounded corners] (1,-1) rectangle (2,-2);
\draw (1.5,1.5) node {$B^n_L$};
\draw (1.5,-1.5) node {$\bar{B}^n_L$};
\draw (2,1.5) -- (2.5,1.5); 
\draw (2,-1.5) -- (2.5,-1.5);
\draw[rounded corners] (2.5,2) rectangle (4.5,-2);
\draw (3.5,0) node {$(\EZL)^m$};
\draw (4.5,1.5) -- (5.0,1.5);  
\draw (4.5,-1.5) -- (5.0,-1.5);
\draw[rounded corners] (5,2) rectangle (6,1);
\draw[rounded corners] (5,-1) rectangle (6,-2);
\draw (1.5,1) -- (1.5,-1);
\draw (5.5,1.5) node {$X_Z$};
\draw (5.5,0) node(X) {};
\draw (5.5,-1.5) node {$\bar{X}_Z$};
\draw (6,-1.5) edge[out=0,in=0] (6,1.5);
\end{tikzpicture} 
\\
&+\sum_{i=1}^{n-1}
\begin{tikzpicture}[baseline = (X.base),every node/.style={scale=0.6},scale=.4]
\draw (1,-1.5) edge[out=180,in=180] (1,1.5);
\draw[rounded corners] (1,2) rectangle (2,1);
\draw[rounded corners] (1,-1) rectangle (2,-2);
\draw (1.5,1.5) node {$X_{B^i}$};
\draw (1.5,-1.5) node {$\bar{X}_{B^i}$};
\draw (2,-1.5) edge[out=0,in=0] (2,1.5);
\end{tikzpicture} 
+
\begin{tikzpicture}[baseline = (X.base),every node/.style={scale=0.6},scale=.4]
\draw (1,-1.5) edge[out=180,in=180] (1,1.5);
\draw[rounded corners] (1,2) rectangle (2,1);
\draw[rounded corners] (1,-1) rectangle (2,-2);
\draw (1.5,1.5) node {$b^n_{L}$};
\draw (1.5,-1.5) node {$\bar{b}^n_{L}$};
\draw (2,-1.5) edge[out=0,in=0] (2,1.5);
\draw (1.5,1) -- (1.5,-1);
\end{tikzpicture}.
\end{split}
\end{equation*}
To simplify $\braket{\Phi|\Phi}$ further, we explicitly split out the contribution of $\ER$ from its leading eigenspace: 
\begin{equation}
  \ER = 
\begin{tikzpicture}[baseline = (X.base),every node/.style={scale=0.60},scale=.4]
\draw (0.5,-1.5) edge[out=0,in=0] (0.5,1.5);
\draw (2.0,0) circle (.5);
\draw (2,0) node {$l_{A_R}$};
\draw (3,1.5) edge[out=180,in=90] (2,0.5);
\draw (3,-1.5) edge[out=180,in=270] (2,-0.5);
\end{tikzpicture} 
+ 
\tilde{E}_{A_R}
\end{equation}
where $l_{A_R}$ is the leading left-eigenvector of $\ER$, and $\tilde{E}_{A_R}$ is the contribution from the sub-leading eigenspace of $\ER$.  
Then, 
\begin{equation}
\sum_{m=0}^\infty
\begin{tikzpicture}[baseline = (X.base),every node/.style={scale=0.6},scale=.4]
\draw (1.5,1.5) -- (2.5,1.5); 
\draw (1.5,-1.5) -- (2.5,-1.5);
\draw[rounded corners] (2.5,2) rectangle (4.5,-2);
\draw (3.5,0) node {$(\ER)^m$};
\draw (4.5,1.5) -- (5.5,1.5);  
\draw (4.5,-1.5) -- (5.5,-1.5);
\end{tikzpicture} 
=
\sum_{m=0}^\infty \begin{tikzpicture}[baseline = (X.base),every node/.style={scale=0.6},scale=.4]
\draw (0.5,-1.5) edge[out=0,in=0] (0.5,1.5);
\draw (2.0,0) circle (.5);
\draw (2,0) node {$l_{A_R}$};
\draw (3,1.5) edge[out=180,in=90] (2,0.5);
\draw (3,-1.5) edge[out=180,in=270] (2,-0.5);
\end{tikzpicture}
+
\sum_{m=0}^\infty
\begin{tikzpicture}[baseline = (X.base),every node/.style={scale=0.6},scale=.4]
\draw (1.5,1.5) -- (2.5,1.5); 
\draw (1.5,-1.5) -- (2.5,-1.5);
\draw[rounded corners] (2.5,2) rectangle (4.5,-2);
\draw (3.5,0) node {$(\tilde{E}_{A_R})^m$};
\draw (4.5,1.5) -- (5.5,1.5);  
\draw (4.5,-1.5) -- (5.5,-1.5);
\end{tikzpicture}. 
\label{eq:split}
\end{equation}
This splitting is useful because $\tilde{E}_{A_R}$ has a spectral radius less than one, and the second term on the right-hand side of Eq. \ref{eq:split} converges.    
We now have
\begin{equation}
\sum_{m=0}^\infty
\begin{tikzpicture}[baseline = (X.base),every node/.style={scale=0.6},scale=.4]
\draw (1,-1.5) edge[out=180,in=180] (1,1.5);
\draw[rounded corners] (1,2) rectangle (2,1);
\draw[rounded corners] (1,-1) rectangle (2,-2);
\draw (1.5,1.5) node {$X_A$};
\draw (1.5,-1.5) node {$\bar{X}_A$};
\draw (2,1.5) -- (2.5,1.5); 
\draw (2,-1.5) -- (2.5,-1.5);
\draw[rounded corners] (2.5,2) rectangle (4.5,-2);
\draw (3.5,0) node {$(\ER)^m$};
\draw (4.5,1.5) -- (5.0,1.5);  
\draw (4.5,-1.5) -- (5.0,-1.5);
\draw[rounded corners] (5,2) rectangle (6,1);
\draw[rounded corners] (5,-1) rectangle (6,-2);
\draw (5.5,1) -- (5.5,-1);
\draw (5.5,1.5) node {$B^1_R$};
\draw (5.5,0) node(X) {};
\draw (5.5,-1.5) node {$\bar{B}^1_R$};
\draw (6,-1.5) edge[out=0,in=0] (6,1.5);
\end{tikzpicture} 
= 
\sum_{m=0}^\infty 
\begin{tikzpicture}[baseline = (X.base),every node/.style={scale=0.6},scale=.4]
\draw (1,-1.5) edge[out=180,in=180] (1,1.5);
\draw[rounded corners] (1,2) rectangle (2,1);
\draw[rounded corners] (1,-1) rectangle (2,-2);
\draw (1.5,1.5) node {$X_A$};
\draw (1.5,-1.5) node {$\bar{X}_A$};
\draw (2,-1.5) edge[out=0,in=0] (2,1.5);
\end{tikzpicture} 
+F_A
\label{eq:FA}
\end{equation} 
where $F_A$ is a finite number. 
Here we have used the normalization of the state: 
\begin{equation}
  \braket{\Psi|\Psi} =  
\begin{tikzpicture}[baseline = (X.base),every node/.style={scale=0.6},scale=.4]
\draw (4.,0) circle (.5);
\draw (4.,0) node {$l_{A_R}$};
\draw (5,1.5) edge[out=180,in=90] (4,0.5);
\draw (5,-1.5) edge[out=180,in=270] (4,-0.5);
\draw[rounded corners] (5,2) rectangle (6,1);
\draw[rounded corners] (5,-1) rectangle (6,-2);
\draw (5.5,1) -- (5.5,-1);
\draw (5.5,1.5) node {$B^1_R$};
\draw (5.5,0) node(X) {};
\draw (5.5,-1.5) node {$\bar{B}^1_R$};
\draw (6,-1.5) edge[out=0,in=0] (6,1.5);
\end{tikzpicture} = 1.
\label{eq:normalization}
\end{equation}
An relation analogous to Eq. \ref{eq:FA} holds for the $Z$ tensors, too. 
This gives the final form of the Gram matrix:
\begin{equation}
  \begin{split}
  \braket{\Phi(&\bar{X}_A;\bar{X}_{B^i};\bar{X}_Z;\bar{b}^{n}_L)|\Phi({X}_A;{X}_{B^i};{X}_Z;{b}_L^n)}
\\
  &= 
\sum_{m=0}^\infty 
\begin{tikzpicture}[baseline = (X.base),every node/.style={scale=0.6},scale=.4]
\draw (1,-1.5) edge[out=180,in=180] (1,1.5);
\draw[rounded corners] (1,2) rectangle (2,1);
\draw[rounded corners] (1,-1) rectangle (2,-2);
\draw (1.5,1.5) node {$X_A$};
\draw (1.5,-1.5) node {$\bar{X}_A$};
\draw (2,-1.5) edge[out=0,in=0] (2,1.5);
\end{tikzpicture} 
+
\sum_{m=0}^\infty 
\begin{tikzpicture}[baseline = (X.base),every node/.style={scale=0.6},scale=.4]
\draw (1,-1.5) edge[out=180,in=180] (1,1.5);
\draw[rounded corners] (1,2) rectangle (2,1);
\draw[rounded corners] (1,-1) rectangle (2,-2);
\draw (1.5,1.5) node {$X_Z$};
\draw (1.5,-1.5) node {$\bar{X}_Z$};
\draw (2,-1.5) edge[out=0,in=0] (2,1.5);
\end{tikzpicture} 
+
\begin{tikzpicture}[baseline = (X.base),every node/.style={scale=0.6},scale=.4]
\draw (1,-1.5) edge[out=180,in=180] (1,1.5);
\draw[rounded corners] (1,2) rectangle (2,1);
\draw[rounded corners] (1,-1) rectangle (2,-2);
\draw (1.5,1.5) node {$b^n_{L}$};
\draw (1.5,-1.5) node {$\bar{b}^n_{L}$};
\draw (2,-1.5) edge[out=0,in=0] (2,1.5);
\draw (1.5,1) -- (1.5,-1);
\end{tikzpicture} 
\\
&+\sum_{i=1}^{n-1}
\begin{tikzpicture}[baseline = (X.base),every node/.style={scale=0.6},scale=.4]
\draw (1,-1.5) edge[out=180,in=180] (1,1.5);
\draw[rounded corners] (1,2) rectangle (2,1);
\draw[rounded corners] (1,-1) rectangle (2,-2);
\draw (1.5,1.5) node {$X_{B^i}$};
\draw (1.5,-1.5) node {$\bar{X}_{B^i}$};
\draw (2,-1.5) edge[out=0,in=0] (2,1.5);
\end{tikzpicture} 
+ F_A(X_A) + F_Z(X_Z).
\end{split}
\label{eq:Gram}
\end{equation}
The Gram matrix is thus essentially diagonal in the effective parameters of $\ket{\Phi}$.

We are ready to compute the orthogonal projection of $\hat{H}\ket{\Psi}$, which is given by the solution to the minimization problem
\begin{equation*}
  \min_{X_A,X_{B^i},X_Z,b^n_L}\norm{\HPsi - \ket{\Phi({X}_A;{X}_{B^i};{X}_Z;{b}^{n}_L)}}_2^2.
\end{equation*}
$X_A$ is determined by   
\begin{equation}
  \frac{\partial\braket{\Phi|\Phi}}{\partial \bar{X}_A}  =  
  \frac{\partial F_A}{\partial \bar{X}_A} + 
  \sum_{m=0}^\infty
\begin{tikzpicture}[baseline = (X.base),every node/.style={scale=0.60},scale=.4]
\draw (0.0,-0.5) -- (0.5,-0.5);
\draw[rounded corners] (0.5,0) rectangle (1.5,-1);  
\draw (1,-0.5) node (X) {$X_A$};
\draw (1.5,-0.5) -- (2.0,-0.5);
\end{tikzpicture} 
= \frac{\partial\braket{\Phi | \hat{H}|\Psi}}{\partial \bar{X}_A}. 
\label{eq:X_A}
\end{equation}
Here,
\begin{equation}
  \begin{split}
   &\frac{\partial\braket{\Phi | \hat{H}|\Psi}}{\partial \bar{X}_A} 
\\ 
& =\sum_{m=0}^\infty \cdots
 \begin{tikzpicture}[baseline = (X.base),every node/.style={scale=0.6},scale=.4]
\draw (2.5,3.5) -- (3,3.5);
\draw (2.5,1.5) -- (3,1.5);
\draw (2.5,-0.5) -- (3,-0.5);
\draw[rounded corners] (3, 4) rectangle (4.2,-1);
\draw (3.6,1.5) node {$\TL$};
\draw (4.2,3.5)  -- (5.2,3.5);
\draw (4.2,1.5)  -- (5.2,1.5);
\draw (4.2,-0.5) -- (5.2,-0.5);
\draw[rounded corners] (5.2,4) rectangle (6.2,3);
\draw (5.7,3.5) node {$A_C$};
\draw (5.7,3.0) -- (5.7,2.0); 
\draw[rounded corners] (5.2,2.0) rectangle (6.2,1);
\draw (5.7,1.5) node {$W_A$};
\draw (5.7,1.0) -- (5.7,-0.); 
\draw[rounded corners] (5.2,0.0) rectangle (6.2,-1);
\draw (5.7,-0.5) node {$\bar{V}_{A_L}$};
\draw (6.2,1.5) -- (7.7,1.5); 
\draw (6.2,3.5) -- (7.7,3.5); 
\draw (6.2,-.5) -- (6.7,-.5); 
edge\draw (6.7,-0.5) edge[out=0,in=0] (6.7,-1.5); 
\draw (7.7,-0.5) edge[out=180,in=180] (7.7,-1.5); 
\draw (7.7,1.5) -- (8.2,1.5); 
\draw (7.7,3.5) -- (8.2,3.5); 
\draw (7.7,-.5) -- (8.2,-.5); 
\draw[rounded corners] (8.2, 4) rectangle (10.2,-1);
\draw (9.2,1.5) node {$(\TAR)^m$};
\draw (10.2,3.5) -- (11.2,3.5);
\draw (10.2,1.5) -- (11.2,1.5);
\draw (10.2,-.5) -- (11.2,-0.5);
\draw[rounded corners] (11.2, 4) rectangle (13.2,-1);
\draw (12.25,1.5) node {$\displaystyle\prod_{i=1}^n E_{B^i_R}^\site{W}$};
\draw (13.2,3.5) -- (14.2,3.5);
\draw (13.2,1.5) -- (14.2,1.5);
\draw (13.2,-.5) -- (14.2,-0.5);
\draw (5,1.5) node (X) {$\phantom{X}$};,
\draw[rounded corners] (14.2, 4) rectangle (15.4,-1);
\draw (14.8,1.5) node {$\TR$};
\draw (15.4,3.5) -- (15.9,3.5);
\draw (15.4,1.5) -- (15.9,1.5);
\draw (15.4,-.5) -- (15.9,-.5);
\end{tikzpicture}\cdots 
\end{split}
\label{eq:long}
\end{equation}
where the MPO transfer matrices $E_{A_L}^\site{W}$, etc., are defined in Eq. \ref{eq:EW}. 
In addition to their generalized eigenvectors, we denote the left eigenvectors of $E_{A_L}^\site{W}$ and right eigenvectors of $E_{A_R}^\site{W}$ respectively as $\rbra{I_1}$ and $\rket{I_{d_W}}$.   
In fact, these eigenvectors do not depend on the values of the MPO, and thus are the same for $E_{Z_L}$ and $E_{Z_R}$ (see Appendix \ref{app:schur}). 
As the left boundary tensor at left infinity propagates through infinitely many $\TL$ to meet the center site $A_C$ in Eq. \ref{eq:long}, only the leading eigenspace survives. 
The same applies to the right side. 
Thus,  
\begin{equation}
  \frac{\partial\braket{\Phi | \hat{H}|\Psi}}{\partial \bar{X}_A}  = \left[\rbra{L_A^\site{W}} + \alpha \rbra{I_1}\right] E_C \left[\rket{R_Z^\site{W}} + \beta \rket{I_{d_W}}\right], 
  \label{eq:E_C}
\end{equation}
where 
\begin{equation}
  E_C \equiv  
 \begin{tikzpicture}[baseline = (X.base),every node/.style={scale=0.6},scale=.4]
\draw (4.7,3.5)  -- (5.2,3.5);
\draw (4.7,1.5)  -- (5.2,1.5);
\draw (4.7,-0.5) -- (5.2,-0.5);
\draw[rounded corners] (5.2,4) rectangle (6.2,3);
\draw (5.7,3.5) node {$A_C$};
\draw (5.7,3.0) -- (5.7,2.0); 
\draw[rounded corners] (5.2,2.0) rectangle (6.2,1);
\draw (5.7,1.5) node {$W_A$};
\draw (5.7,1.0) -- (5.7,-0.); 
\draw[rounded corners] (5.2,0.0) rectangle (6.2,-1);
\draw (5.7,-0.5) node {$\bar{V}_{A_L}$};
\draw (6.2,1.5) -- (7.7,1.5); 
\draw (6.2,3.5) -- (7.7,3.5); 
\draw (6.2,-.5) -- (6.7,-.5); 
edge\draw (6.7,-0.5) edge[out=0,in=0] (6.7,-1.5); 
\draw (7.7,-0.5) edge[out=180,in=180] (7.7,-1.5); 
\draw (7.7,1.5) -- (8.2,1.5); 
\draw (7.7,3.5) -- (8.2,3.5); 
\draw (7.7,-.5) -- (8.2,-.5); 
\draw[rounded corners] (8.2, 4) rectangle (10.2,-1);
\draw (9.2,1.5) node {$(\TAR)^m$};
\draw (10.2,3.5) -- (11.2,3.5);
\draw (10.2,1.5) -- (11.2,1.5);
\draw (10.2,-.5) -- (11.2,-0.5);
\draw[rounded corners] (11.2, 4) rectangle (13.2,-1);
\draw (12.25,1.5) node {$\displaystyle\prod_{i=1}^n E_{B^i_R}^\site{W}$};
\draw (13.2,3.5) -- (13.7,3.5);
\draw (13.2,1.5) -- (13.7,1.5);
\draw (13.2,-.5) -- (13.7,-0.5);
\draw (5,1.5) node (X) {$\phantom{X}$};,
\end{tikzpicture}.  
\end{equation}
Here, $\alpha$ and $\beta$ are two complex numbers. 
They occur because every time $\rbra{L_A^\site{W}}$ passes through $\TL$, there arises a new term of $\rbra{I_1}$: $ \rbra{L_A^\site{W}} \TL = \rbra{L_A^\site{W}} + e \rbra{I_1}$, where $e$ is the energy density of the chain \cite{VUMPS}.
Their values, however, do not matter because of the following lemmas.
\begin{lemma}
$\rbra{I_1}E_C = 0$. 
(This lemma, and others below, are based on the Schur form of the MPO.  
See Appendix \ref{app:schur} for a discussion of their proofs.)
\label{lem:1}
\end{lemma}
\begin{lemma}
$\rbra{L_A^\site{W}}E_C\rket{I_{d_W}} = 0$. 
\label{lem:2}
\end{lemma}
Thus, 
\begin{equation*}
  \begin{split}
\frac{\partial\braket{\Phi | \hat{H}|\Psi}}{\partial \bar{X}_A}
& =\sum_{m=0}^\infty
 \begin{tikzpicture}[baseline = (X.base),every node/.style={scale=0.6},scale=.4]
\draw[rounded corners] (3, 4) rectangle (4.2,-1);
\draw (3.6,1.5) node {$L_A^\site{W}$};
\draw (4.2,3.5)  -- (5.2,3.5);
\draw (4.2,1.5)  -- (5.2,1.5);
\draw (4.2,-0.5) -- (5.2,-0.5);
\draw[rounded corners] (5.2,4) rectangle (6.2,3);
\draw (5.7,3.5) node {$A_C$};
\draw (5.7,3.0) -- (5.7,2.0); 
\draw[rounded corners] (5.2,2.0) rectangle (6.2,1);
\draw (5.7,1.5) node {$W_A$};
\draw (5.7,1.0) -- (5.7,-0.); 
\draw[rounded corners] (5.2,0.0) rectangle (6.2,-1);
\draw (5.7,-0.5) node {$\bar{V}_{A_L}$};
\draw (6.2,1.5) -- (7.7,1.5); 
\draw (6.2,3.5) -- (7.7,3.5); 
\draw (6.2,-.5) -- (6.7,-.5); 
edge\draw (6.7,-0.5) edge[out=0,in=0] (6.7,-1.5); 
\draw (7.7,-0.5) edge[out=180,in=180] (7.7,-1.5); 
\draw (7.7,1.5) -- (8.2,1.5); 
\draw (7.7,3.5) -- (8.2,3.5); 
\draw (7.7,-.5) -- (8.2,-.5); 
\draw[rounded corners] (8.2, 4) rectangle (10.2,-1);
\draw (9.2,1.5) node {$(\TAR)^m$};
\draw (10.2,3.5) -- (11.2,3.5);
\draw (10.2,1.5) -- (11.2,1.5);
\draw (10.2,-.5) -- (11.2,-0.5);
\draw[rounded corners] (11.2, 4) rectangle (13.2,-1);
\draw (12.25,1.5) node {$\displaystyle\prod_{i=1}^n E_{B^i_R}^\site{W}$};
\draw (13.2,3.5) -- (14.2,3.5);
\draw (13.2,1.5) -- (14.2,1.5);
\draw (13.2,-.5) -- (14.2,-0.5);
\draw (5,1.5) node (X) {$\phantom{X}$};,
\draw[rounded corners] (14.2, 4) rectangle (15.4,-1);
\draw (14.8,1.5) node {$R_Z^\site{W}$};
\end{tikzpicture}. 
\end{split}
\end{equation*}
As with $E_{A_R}$, we split out of $\TAR$ the term associated with the leading eigenspace. 
To do this, we need the following lemma in linear algebra. 
\begin{lemma}
Let $E$ be a matrix with leading eigenvalue one, according to which there is one eigenvector and one generalized eigenvector.  
Let $\rbra{v_1}$ be the left generalized eigenvector, $\rbra{v_2}$ the left eigenvector, $\rket{u_1}$ the right eigenvector, and $\rket{u_2}$ the right generalized eigenvector.  
Then, for an integer $m > 0$,  
\begin{equation}
  E^m =  \rket{u_1}\rbra{v_1} + m \rket{u_1}\rbra{v_2} + \rket{u_2}\rbra{v_2} + \tilde{E}^m, 
\end{equation}
where $\tilde{E}$ is the contribution to $E$ from the sub-leading eigenspace. 
\label{lem:3}
\end{lemma}
When applying Lemma \ref{lem:3} to $\TAR$, the contribution associated with the $\rket{u_1} = \rket{I_{d_W}}$ drops because of the following lemma.  
\begin{lemma}
  \begin{equation}
 \begin{tikzpicture}[baseline = (X.base),every node/.style={scale=0.6},scale=.4]
\draw[rounded corners] (3, 4) rectangle (4.2,-1);
\draw (3.6,1.5) node {$L_A^\site{W}$};
\draw (4.2,3.5)  -- (5.2,3.5);
\draw (4.2,1.5)  -- (5.2,1.5);
\draw (4.2,-0.5) -- (5.2,-0.5);
\draw[rounded corners] (5.2,4) rectangle (6.2,3);
\draw (5.7,3.5) node {$A_C$};
\draw (5.7,3.0) -- (5.7,2.0); 
\draw[rounded corners] (5.2,2.0) rectangle (6.2,1);
\draw (5.7,1.5) node {$W_A$};
\draw (5.7,1.0) -- (5.7,-0.); 
\draw[rounded corners] (5.2,0.0) rectangle (6.2,-1);
\draw (5.7,-0.5) node {$\bar{V}_{A_L}$};
\draw (6.2,1.5) -- (7.7,1.5); 
\draw (6.2,3.5) -- (7.7,3.5); 
\draw (6.2,-.5) -- (6.7,-.5); 
edge\draw (6.7,-0.5) edge[out=0,in=0] (6.7,-1.5); 
\draw (7.7,-0.5) edge[out=180,in=180] (7.7,-1.5); 
\draw (7.7,1.5) -- (8.2,1.5); 
\draw (7.7,3.5) -- (8.2,3.5); 
\draw (7.7,-.5) -- (8.2,-.5); 
\draw[rounded corners] (8.2, 4) rectangle (9.4,-1);
\draw (8.8,1.5) node {$I_{d_W}$};
\end{tikzpicture} = 0. 
  \end{equation}
  \label{lem:4}
\end{lemma}
Thus, we have 
\begin{equation}
  \begin{split}
&\frac{\partial\braket{\Phi | \hat{H}|\Psi}}{\partial \bar{X}_A}
 =\sum_{m=0}^\infty
 \begin{tikzpicture}[baseline = (X.base),every node/.style={scale=0.6},scale=.4]
\draw[rounded corners] (3, 4) rectangle (4.2,-1);
\draw (3.6,1.5) node {$L_A^\site{W}$};
\draw (4.2,3.5)  -- (5.2,3.5);
\draw (4.2,1.5)  -- (5.2,1.5);
\draw (4.2,-0.5) -- (5.2,-0.5);
\draw[rounded corners] (5.2,4) rectangle (6.2,3);
\draw (5.7,3.5) node {$A_C$};
\draw (5.7,3.0) -- (5.7,2.0); 
\draw[rounded corners] (5.2,2.0) rectangle (6.2,1);
\draw (5.7,1.5) node {$W_A$};
\draw (5.7,1.0) -- (5.7,-0.); 
\draw[rounded corners] (5.2,0.0) rectangle (6.2,-1);
\draw (5.7,-0.5) node {$\bar{V}_{A_L}$};
\draw (6.2,1.5) -- (7.7,1.5); 
\draw (6.2,3.5) -- (7.7,3.5); 
\draw (6.2,-.5) -- (6.7,-.5); 
edge\draw (6.7,-0.5) edge[out=0,in=0] (6.7,-1.5); 
\draw (7.7,-0.5) edge[out=180,in=180] (7.7,-1.5); 
\draw (7.7,1.5) -- (8.2,1.5); 
\draw (7.7,3.5) -- (8.2,3.5); 
\draw (7.7,-.5) -- (8.2,-.5); 
\draw[rounded corners] (8.2, 4) rectangle (9.4,-1);
\draw (8.8,1.5) node {$R_{A}^\site{W}$};
\end{tikzpicture} 
\\ 
& \hspace{5mm}+
 \begin{tikzpicture}[baseline = (X.base),every node/.style={scale=0.6},scale=.4]
\draw[rounded corners] (3, 4) rectangle (4.2,-1);
\draw (3.6,1.5) node {$L_A^\site{W}$};
\draw (4.2,3.5)  -- (5.2,3.5);
\draw (4.2,1.5)  -- (5.2,1.5);
\draw (4.2,-0.5) -- (5.2,-0.5);
\draw[rounded corners] (5.2,4) rectangle (6.2,3);
\draw (5.7,3.5) node {$A_C$};
\draw (5.7,3.0) -- (5.7,2.0); 
\draw[rounded corners] (5.2,2.0) rectangle (6.2,1);
\draw (5.7,1.5) node {$W_A$};
\draw (5.7,1.0) -- (5.7,-0.); 
\draw[rounded corners] (5.2,0.0) rectangle (6.2,-1);
\draw (5.7,-0.5) node {$\bar{V}_{A_L}$};
\draw (6.2,1.5) -- (7.7,1.5); 
\draw (6.2,3.5) -- (7.7,3.5); 
\draw (6.2,-.5) -- (6.7,-.5); 
edge\draw (6.7,-0.5) edge[out=0,in=0] (6.7,-1.5); 
\draw (7.7,-0.5) edge[out=180,in=180] (7.7,-1.5); 
\draw (7.7,1.5) -- (8.2,1.5); 
\draw (7.7,3.5) -- (8.2,3.5); 
\draw (7.7,-.5) -- (8.2,-.5); 
\draw[rounded corners] (8.2, 4) rectangle (11.2,-1);
\draw (9.7,1.5) node {$\displaystyle\sum_{m=0}^\infty (\tilde{E}^{[W]}_{A_R})^m$};
\draw (11.2,3.5) -- (12.2,3.5);
\draw (11.2,1.5) -- (12.2,1.5);
\draw (11.2,-.5) -- (12.2,-0.5);
\draw[rounded corners] (12.2, 4) rectangle (14.2,-1);
\draw (13.25,1.5) node {$\displaystyle\prod_{i=1}^n E_{B^i_R}^\site{W}$};
\draw (14.2,3.5) -- (15.2,3.5);
\draw (14.2,1.5) -- (15.2,1.5);
\draw (14.2,-.5) -- (15.2,-0.5);
\draw (5,1.5) node (X) {$\phantom{X}$};,
\draw[rounded corners] (15.2, 4) rectangle (16.4,-1);
\draw (15.8,1.5) node {$R_Z^\site{W}$};
\end{tikzpicture}, 
\end{split}
\label{eq:jj}
\end{equation}
where we have made use of the following lemma. 
\begin{lemma}
  \begin{equation}
 \begin{tikzpicture}[baseline = (X.base),every node/.style={scale=0.6},scale=.4]
\draw[rounded corners] (3, 4) rectangle (4.2,-1);
\draw (3.6,1.5) node {$l_{A_R}^\site{W}$};
\draw (4.2,3.5)  -- (5.2,3.5);
\draw (4.2,1.5)  -- (5.2,1.5);
\draw (4.2,-0.5) -- (5.2,-0.5);
\draw[rounded corners] (5.2, 4) rectangle (7.2,-1);
\draw (6.25,1.5) node {$\displaystyle\prod_{i=1}^n E_{B^i_R}^\site{W}$};
\draw (7.2,3.5)  -- (8.2,3.5);
\draw (7.2,1.5)  -- (8.2,1.5);
\draw (7.2,-0.5) -- (8.2,-0.5);
\draw[rounded corners] (8.2, 4) rectangle (9.4,-1);
\draw (8.8,1.5) node {$R_{Z}^\site{W}$};
\end{tikzpicture} = 1, 
  \end{equation}
where $l_{A_R}^\site{W}$ is the left eigenvector of $\TAR$. 
\label{lem:5}
\end{lemma}

Note that the second term of Eq. \ref{eq:jj} converges. 
Now substitute Eq. \ref{eq:jj} into Eq. \ref{eq:X_A}, and divide the equation by $\sum_{m=0}^\infty 1$. 
The finite terms drop, and we obtain Eq. \ref{eq:X_A_final}. 
Analogously, we obtain Eq. \ref{eq:X_Z_final} and \ref{eq:X_B_final}.    

We now determine $b_L^n$, which is given by 
\begin{equation*}
\begin{split}
\frac{\partial\braket{\Phi|\Phi}}{\partial \bar{b}_L^n}  &=   
\begin{tikzpicture}[baseline = (X.base),every node/.style={scale=0.60},scale=.4]
\draw (0,-0.5) -- (0.5,-0.5);
\draw[rounded corners] (0.5,0) rectangle (1.5,-1);  
\draw (1,-0.5) node {$b_L^n$};
\draw (1,-0.8) node(X) {};
\draw (1,-1) -- (1,-1.5);
\draw (1.5,-0.5) -- (2,-0.5);
\end{tikzpicture} 
= \frac{\partial\braket{\Phi | \hat{H}|\Psi}}{\partial \bar{b}_L^n} 
\\
&=
  \left[\rbra{L_A^\site{W}} + \alpha \rbra{I_1}\right] 
  E_D
\left[\rket{R_Z^\site{W}} + \beta \rket{I_{d_W}}\right],
\end{split}
\label{eq:b_n}
\end{equation*}
where 
\begin{equation}
  E_D \equiv 
\begin{tikzpicture}[baseline = (X.base),every node/.style={scale=0.6},scale=.4]
\draw (5,1.5) node (X) {$\phantom{X}$};,
\draw (4.2,3.5)  -- (5.2,3.5);
\draw (4.2,1.5)  -- (5.2,1.5);
\draw (4.2,-0.5) -- (5.2,-0.5);
\draw[rounded corners] (5.2, 4) rectangle (7.2,-1);
\draw (6.25,1.5) node {$\displaystyle\prod_{i=1}^{n-1} E_{B^i_L}^\site{W}$};
\draw (7.2,3.5) -- (8.2,3.5);
\draw (7.2,1.5) -- (8.2,1.5);
\draw (7.2,-.5) -- (7.7,-0.5);
edge\draw (7.7,-0.5) edge[out=0,in=0] (7.7,-1.5); 
\draw[rounded corners] (8.2,4) rectangle (9.2,3);
\draw (8.7,3.5) node {$B^n_C$};
\draw (8.7,3.0) -- (8.7,2.0); 
\draw[rounded corners] (8.2,2.0) rectangle (9.2,1);
\draw (8.7,1.5) node {$W_n$};
\draw (8.7,1.0) -- (8.7,-1.5); 
\draw (9.2,3.5) -- (10.2,3.5);
\draw (9.2,1.5) -- (10.2,1.5);
\draw (9.7,-.5) -- (10.2,-0.5);
\draw (9.7,-0.5) edge[out=180,in=180] (9.7,-1.5); 
\end{tikzpicture}. 
\end{equation}
Here the $\alpha$ and $\beta$ are the same as in Eq. \ref{eq:E_C}. 
Two lemmas are now in order: 
\begin{lemma}
  $\rbra{I_1}E_D\rket{I_{d_W}} = 0$. 
  \label{lem:6}
\end{lemma}
\begin{lemma}
$  \rbra{I_1}E_D\rket{R_{Z}^\site{W}} =  
\rbra{L_A^\site{W}}E_D\rket{I_{d_W}} = B_C^n. 
$
\label{lem:7}
\end{lemma}
Thus, 
\begin{equation}
  b_L^n = 
  \rbra{L_A^\site{W}} E_D \rket{R_Z^\site{W}} + (\alpha+\beta)B_C^n. 
  \label{eq:bn_L}
\end{equation}
But note that $b^n_L = (\alpha+\beta)B_C^n$ gives a contribution of $(\alpha+\beta)\ket{\Psi}$ to $\ket{\Phi}$, which can be dropped in the projective space.  
Also recall that we still have one last gauge symmetry to spare, which we now use to demand $\alpha + \beta = 0$ so that $b_L^n = \rbra{L_A^\site{W}} E_D \rket{R_Z^\site{W}}$ in Eq. \ref{eq:bn_final}. 
\subsection{Schur form of MPO}
\label{app:schur}
As discussed in the main text, the $W$ matrix of an MPO is lower-triangular, known as the Schur form. 
For example, in terms of the operator-valued matrices $\hat{W}_{ab} = \sum_{ss'}W^{ss'}_{ab}\ket{s}\bra{s'}$, the $W$ matrix of the transverse-field Ising Hamiltonian (when $h_z = 0$) in Eq. \ref{eq:Ising} can be expressed as, 
\begin{equation}
 \hat{W}=
 \begin{bmatrix}
\unity&0&0\\
-\hat \sigma^z&0&0\\
h_x\hat \sigma^x&\hat \sigma^z&\unity
\end{bmatrix}
\label{eq:TFI_MPO}
\end{equation} 
where $\hat \sigma^x$ and $\hat \sigma^z$ are the Pauli matrices. 
To us, the important features of $\hat{W}$ are that $\hat{W}$ is lower triangular and that $\hat{W}_{11} = \hat{W}_{d_Wd_W} = \unity$.   
This means that the dominant left-eigenvector $\rbra{I_1}$ of $\TL$ and right-eigenvector $\rket{I_{d_W}}$ of $\TR$ are 
\begin{equation}
\begin{tikzpicture}[baseline = (X.base),every node/.style={scale=0.6},scale=.4]
\draw[rounded corners] (3,4) rectangle (4,-1);
\draw (3.5,1.5) node[scale=1.5] {$I_1$};
\draw (4,-0.5) -- (5,-0.5); 
\draw (4,1.5) -- (5,1.5);
\draw (5.5,1.5) node[scale=1.5] {$a$};
\draw (4,3.5) -- (5,3.5);
\draw (5,1.5) node (X) {$\phantom{X}$};
\end{tikzpicture} 
= 
\delta_{a1}
\begin{tikzpicture}[baseline = (X.base),every node/.style={scale=0.6},scale=.4]
\draw (1,-1) edge[out=180,in=180] (1,4);
\end{tikzpicture},
\hspace{8mm}
\begin{tikzpicture}[baseline = (X.base),every node/.style={scale=0.6},scale=.4]
\draw[rounded corners] (5,4) rectangle (6.5,-1);
\draw (5.75,1.5) node[scale=1.5] {$I_{d_W}$};
\draw (4,-0.5) -- (5,-0.5); 
\draw (4,1.5) -- (5,1.5);
\draw (3.5,1.5) node[scale=1.5] {$a$};
\draw (4,3.5) -- (5,3.5);
\draw (5,1.5) node (X) {$\phantom{X}$};
\end{tikzpicture} 
= 
\delta_{ad_W}
\begin{tikzpicture}[baseline = (X.base),every node/.style={scale=0.6},scale=.4]
\draw (1,-1) edge[out=0,in=0] (1,4);
\end{tikzpicture}.  
\end{equation}
In addition, the generalized eigenvector $\rbra{L_A^\site{W}}$ and $\rket{R_Z^\site{W}}$ satisfy the following relation \cite{VUMPS}:  
\begin{equation}
\begin{tikzpicture}[baseline = (X.base),every node/.style={scale=0.6},scale=.4]
\draw[rounded corners] (2,4) rectangle (4,-1);
\draw (3.0,1.5) node[scale=1.5] {$L^\site{W}_{A}$};
\draw (4,-0.5) -- (5,-0.5); 
\draw (4,1.5) -- (5,1.5);
\draw (5.7,1.5) node[scale=1.5] {$d_W$};
\draw (4,3.5) -- (5,3.5);
\draw (5,1.5) node (X) {$\phantom{X}$};
\end{tikzpicture} 
= 
\begin{tikzpicture}[baseline = (X.base),every node/.style={scale=0.6},scale=.4]
\draw (1,-1) edge[out=180,in=180] (1,4);
\end{tikzpicture},
\hspace{8mm}
\begin{tikzpicture}[baseline = (X.base),every node/.style={scale=0.6},scale=.4]
\draw[rounded corners] (5,4) rectangle (7,-1);
\draw (6,1.5) node[scale=1.5] {$R^\site{W}_{Z}$};
\draw (4,-0.5) -- (5,-0.5); 
\draw (4,1.5) -- (5,1.5);
\draw (3.5,1.5) node[scale=1.5] {$1$};
\draw (4,3.5) -- (5,3.5);
\draw (5,1.5) node (X) {$\phantom{X}$};
\end{tikzpicture} 
= 
\begin{tikzpicture}[baseline = (X.base),every node/.style={scale=0.6},scale=.4]
\draw (1,-1) edge[out=0,in=0] (1,4);
\end{tikzpicture}.  
\end{equation}
We now discuss the proofs of the lemmas in Appendix \ref{app:ortho}. 

Lemma \ref{lem:1}: Because $\rbra{I_1}$ is non-zero only when its middle index is one, $W_A$ only contributes a $\unity$ to $\rbra{I_1}E_C$. 
Thus, $\rbra{I_1}E_C = 0$ by Eq. \ref{eq:V}.   

Lemma \ref{lem:2}: Because $\rket{I_{d_W}}$ is non-zero only when its middle index is $d_W$, and that the only non-zero element in the $d_W$ column of $W$ is $W_{d_Wd_W}$, the $\rbra{L_A^\site{W}}$ contributes only as $\rbra{I_{d_W}}$.     
This makes $\rbra{L_A^\site{W}}E_C\rket{I_{d_W}} = 0$ by Eq. \ref{eq:V}. 

Lemma \ref{lem:3}: This is proved by putting $E$ into its Jordan canonical form.  

Lemma \ref{lem:4}: Similar to Lemma \ref{lem:2}.  

Lemma \ref{lem:5}: Because of the Schur form, $\rbra{l_{A_R}^\site{W}}$ is non-zero only when its middle index is 1, and is equal to $\rbra{l_{A_R}}$ in that case.   
Then this lemma reduces to Eq. \ref{eq:normalization}.  

Lemma \ref{lem:6}: Because of the Schur form, $\rbra{I_1}E_D$ is only non-zero when its middle index is 1, but $\rket{I_{d_W}}$ is only non-zero when its middle index is $d_W$.
This makes the whole thing zero. 

Lemma \ref{lem:7}: Similar to Lemma \ref{lem:2}, $\rbra{L_A^\site{W}} (\rket{R_Z^\site{W}})$ contributes only as $\rbra{I_{d_W}} (\rket{I_1})$ and $W_n$ contributes only as $\unity$. 
Thus, the whole expression reduces to $B_C^n$. 
\subsection{Symplectic derivation of TDVP}
\label{app:symp}
The derivations \cite{tdvp} of TDVP in the literature have been based on a variational principle, hence the name.  
This has the benefit of not needing differential geometry, but buries the symplectic structure of TDVP under the heavy calculations in the derivation.   
Here we give a derivation directly from symplectic geometry, which is quite elegant and may be preferable to a person who knows some basic differential geometry. 
We assume knowledge of basic differential geometry at the level of chapter 5 and 8 of \cite{Geometry}. 

Let $\HH$ be a complex vector space with (complex) dimension $m$. 
$\HH$ can also be viewed as a real manifold with real dimension $2m$, and thus with a tangent space $T_\Psi \HH$ at $\Psi \in \HH$ of real dimension $2m$.     
$T_\Psi \HH$ can be complexified to give $(T_\Psi \HH)^\C$ which has complex dimension $2m$. 
Let $J$ be a linear complex structure on $(T_\Psi \HH)^\C$.  
$J^2=1$ and have two eigenvalues $i$ and $-i$, each with an eigenspace of complex dimension $m$. 
$(T_\Psi\HH)^\C$ can then be written as a direct sum of the eigenspaces of $J$: $(T_\Psi\HH)^\C = (T_\Psi \HH)^+ \oplus(T_\Psi \HH)^-$, where $J(T_\Psi \HH)^+ = i(T_\Psi \HH)^+$ and $J(T_\Psi \HH)^-=-i(T_\Psi \HH)^-$. 
Note that $\dim_\C(T_\Psi \HH)^+ = m = \dim_\C \HH$, and a linear isomorphism can be established: $(T_\Psi \HH)^+ \cong \HH$. 
This allows one to extend the inner product of $\HH$ to $(T_\Psi \HH)^+$: 
\begin{equation}
  I(X, Y) \equiv \braket{X|Y}, \hspace{5mm} \forall X,Y \in (T_\Psi \HH)^+ \cong \HH.   
\end{equation}
Note that we do not define an inner product on $(T_\Psi \HH)^-$. 
$I$ allows a definition of a metric $g$ on $(T_\Psi \HH)^\C$:  $\forall X,Y \in (T_\Psi \HH)^+$, 
\begin{equation}
  \begin{split}
  g(\bar{Y}, X) &= I(Y, X), 
  \\
  g(Y, X) &= 0,
  \\
  g(\bar{Y}, \bar{X}) &= 0. 
\end{split}
\end{equation}
This $g$ is known as the Hermitian metric. 
It is such that $g(JX, JY) = g(X, Y)$ for all $X, Y \in (T_\Psi \HH)^\C$. 
$g$ defines a two-form $\Omega$:  
\begin{equation}
  \Omega(X, Y) = g(JX, Y), \hspace{5mm} \forall X,Y \in (T_\Psi \HH)^\C .
\end{equation}
(It is not hard to show $\Omega(X, Y) = -\Omega(Y,X)$.)  
Because vector spaces are ``flat'', $g$ does not change from point to point, thus $d\Omega = 0$.  
This means $\Omega$ is symplectic. 
A manifold with a compatible complex structure $J$, Hermitian structure $I$, Riemannian structure $g$, and symplectic structure $\Omega$ is known as a K{\"a}hler manifold. 
We have essentially shown that any complex vector space with an inner product is K{\"a}hler.

Let $\xi, \eta, \chi, \phi \in (T_\Psi \HH)^+$. 
$\Omega$ and $I$ are connected by the following: 
\begin{equation}
  \begin{split}
  \Omega(\chi + \bar\phi, \xi + \bar\eta) &= g(J(\chi + \bar\phi), \xi+\bar\eta)
  \\
  &= g(i\chi-i\bar\phi, \xi+\bar\eta)
  \\
  &=g(i\chi,\bar\eta) + g(-i\bar\phi,\xi)
  \\
  &=I(\eta,i\chi) + I(i\phi,\xi).
\end{split}
\end{equation}

On $\HH$, for a Hamiltonian operator $\hat{H}$, consider the Hamiltonian flow of the Hamiltonian function $H: \Psi \in \HH \mapsto \braket{\Psi|\hat{H}|\Psi}$. 
For $\xi, \eta$ infinitesimal:  
\begin{equation}
  \begin{split}
  dH(\xi+\bar\eta)|_\Psi &= \braket{\Psi+\eta|\hat{H}|\Psi+\chi} -  \braket{\Psi|\hat{H}|\Psi}  
  \\
  &= I(\eta, \hat{H}\Psi) + I(\Psi, \hat{H} \xi)    
  \\
  & = I(\eta, \hat{H}\Psi) + I(\hat{H}\Psi, \xi) 
  \\
  &= \Omega(X_H, \xi+\bar\eta)
\end{split}
\end{equation}
where $X_H$ is the Hamiltonian flow of $H$:  
\begin{equation}
  X_H = -i\hat{H} \Psi + \overline{-i\hat{H}\Psi}.  
\end{equation}
This is nothing but the Schr{\"o}dinger flow.  
Thus, the Schr{\"o}dinger dynamics can be viewed as a symplectic flow of the Hamiltonian function $H(\Psi)$.  

Now let $M$ be a submanifold of $\HH$. Does $H$ induce a symplectic Schr{\"o}dinger flow on $M$? Yes!
Let the inclusion function from $M$ to $\HH$ be denoted as 
\begin{equation}
  \inc: M \to \HH, \hspace{5mm} \inc: \Psi\in M \mapsto \Psi \in \HH. 
\end{equation}
Both the Hamiltonian function and the symplectic form have a restriction on $M$:  
\begin{equation}
  H_M = H\circ \inc, \hspace{5mm} \Omega_M = \inc^* \Omega. 
\end{equation}
Because the exterior differentiation $d$ and the pullback $\inc^*$ commutes, $d\Omega_M = 0$, and thus $M$ is also symplectic.  
We now look for the Hamiltonian flow $X_{H_M}$ associated with $H_M$ on $M$. 
For all $\xi, \eta \in (T_\Psi M)^+$, we look for $X_{H_M} \in (T_\Psi M)^\C$ such that $\Omega_M(X_{H_M}, \xi+\bar\eta) = dH_M(\xi+\bar\eta)|_\Psi$. 
\begin{equation}
  \begin{split}
    dH_M(\xi+\bar\eta)|_\Psi &= dH(\inc_*(\xi+\bar\eta))|_\Psi
    \\
    &= dH(\xi+\bar\eta)|_\Psi 
    \\
    &= I(\eta, \hat{H}\Psi) + I(\hat{H}\Psi, \xi). 
\end{split}
\end{equation}
Now here is the key, because $\xi, \eta$ are both only in $(T_\Psi M
)^+$, $\hat{H}\Psi$ can be replaced with its orthogonal projection on $(T_\Psi M)^\C$, $\text{Proj} \hat{H}\Psi$:  
\begin{equation}
  \begin{split}
  dH_M(\xi+\bar\eta) & = I(\eta, \text{Proj}\hat{H}\Psi) + I(\text{Proj}\hat{H}\Psi, \xi) 
  \\
  &= \Omega(X_{H_M}, \xi+\bar\eta)
\end{split}
\end{equation}
where $X_{H_M}$ is the Hamiltonian flow of $H_M$ on $M$:  
\begin{equation}
  X_{H_M} = -i\text{Proj}\hat{H} \Psi + \overline{-i\text{Proj}\hat{H}\Psi}.  
\end{equation}
This gives the TDVP dynamics on $M$ and the dynamics is symplectic.  

\bibliography{abc}
\end{document}